\documentclass[draftclsnofoot]{IEEEtran}
 \onecolumn
\usepackage{setspace}
\usepackage{latexsym}
\usepackage{cite}
\usepackage{amssymb}
\usepackage{bm}
\usepackage{amsmath, amssymb}
\usepackage[dvips]{graphics}
\usepackage{graphicx}
\usepackage{psfrag}
\begin{document}
\title{Capacity of the Trapdoor Channel with Feedback}
%\end{titlepage}
%\setlength{\textwidth}{7in} \setlength{\textheight}{9in}
%\setlength{\topmargin}{-0.4in} \setlength{\oddsidemargin}{-0.30in}
%\thispagestyle{empty} \setcounter{page}{1}
%\setlength{\baselineskip}{1\baselineskip} \maketitle
%\IEEEpeerreviewmaketitle
\author{Haim Permuter, Paul Cuff, Benjamin Van Roy and Tsachy Weissman \\
%Information Systems Laboratory\\
%Department of Electrical Engineering\\
%Stanford University\\
%Stanford, CA, 94305 USA \\
%Email: \{haim1, pcuff, bvr, tsachy\}@stanford.edu
%\markboth{Journal
%of \LaTeX\ Class Files,~Vol.~1, No.~11,~November~2002}{Shell
%\MakeLowercase{\textit{et al.}}: Bare Demo of IEEEtran.cls for
%Journals}
\thanks{This work was
supported by National Science Foundation (NSF) through the grants
CCR-0311633, CCF-0515303, IIS-0428868 and the NSF CAREER grant.}
\thanks{The authors are with the Department of Electrical Engineering, Stanford University, Stanford, CA 94305, USA.
(Email: \{haim1, pcuff, bvr, tsachy\}@stanford.edu)}
%\thanks{M. Shell is with the Georgia Institute of Technology.}}
}

\maketitle

%\documentclass[epss,  11pt]{article}
%%\documentclass[a4paper, 10pt, conference]{ieeeconf}
%%\documentclass{IEEEtrans2e}
% \usepackage{setspace}
%\setstretch{1}
%
%%\IEEEoverridecommandlockouts
%%\overrideIEEEmargins
%%\documentclass[epss, 10pt]{article}
%% \input{psfig.sty}
%%\usepackage[notref, notcite]{showkeys}
%%       CHOOSE A HEBREW ENCODING !!!
%%Use the following line for Windows Hebrew
%%\usepackage[cp1255]{inputenc}
%%Use the following line for DOS Hebrew
%%\usepackage[cp862]{inputenc}
%%
%\usepackage{latexsym}
%\usepackage{cite}
%\usepackage{amssymb}
%\usepackage{bm}
%\usepackage{amsmath, amsthm, amssymb}
%%\usepackage{latexsym}
%%\usepackage{amssymb}
%%\usepackage[english,hebrew]{babel}
%\usepackage[dvips]{graphics}
%\usepackage{graphicx}
%\usepackage{psfrag}
%\usepackage{color}
%%\usepackage[english,hebrew]{babel}
%%\usepackage[dvips]{graphics}
%%\pagestyle{headings}
%\pagestyle{plain} \pagenumbering{arabic}
%\begin{titlepage}
%%\begin{document}
%\title{Capacity of the Trapdoor Channel with Feedback}%We'll see
% %\footnote{Both authors are with the Department of
%%Electrical Engineering, Technion- Israel Institute of Technology,
%%Haifa 32000, Israel. tsachy@tx.technion.ac.il,
%%merhav@ee.technion.ac.il.}
%%}
%%\author{Tsachy
%% Weissman}
%\end{titlepage}
%\setlength{\textwidth}{7in} \setlength{\textheight}{9in}
%\setlength{\topmargin}{-0.4in} \setlength{\oddsidemargin}{-0.30in}
%\begin{document}
%\thispagestyle{empty} \setcounter{page}{1}
%\setlength{\baselineskip}{1\baselineskip} \maketitle

\begin{abstract}

We establish that the feedback capacity of the trapdoor channel is
the logarithm of the golden ratio and provide a simple
communication scheme that achieves capacity.  As part of the
analysis, we formulate a class of dynamic programs that
characterize capacities of unifilar finite-state channels.  The
trapdoor channel is an instance that admits a simple analytic
solution.
\end{abstract}


\newtheorem{question}{Question}
\newtheorem{claim}{Claim}
\newtheorem{guess}{Conjecture}
\newtheorem{definition}{Definition}
\newtheorem{fact}{Fact}
\newtheorem{assumption}{Assumption}
\newtheorem{theorem}{Theorem}
\newtheorem{lemma}[theorem]{Lemma}
\newtheorem{ctheorem}{Corrected Theorem}
\newtheorem{corollary}[theorem]{Corollary}
\newtheorem{proposition}{Proposition}
\newtheorem{example}{Example}
\newtheorem{pfth}{Proof}

%\section{Directed Information}

\section{Introduction}

\begin{figure}[h]{
\centerline{\includegraphics[width=9cm]{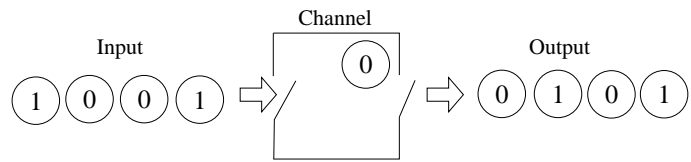}}
%\centerline{\includegraphics[width=9cm]{C:/cygwin/usr/X11R6/bin/home/chemical_horizontal.eps}}
\caption{The trapdoor(chemical) channel.}
\label{f_chemical_horizontal} }
\end{figure}

David Blackwell, who has done fundamental work both in information
theory and in stochastic dynamic programming, introduced the
trapdoor channel in 1961 \cite{Blackwell61} as a ``simple
two-state channel''.  The channel is depicted in Figure
\ref{f_chemical_horizontal}, and a detailed discussion of this
channel appears in the information theory book by Ash
\cite{Ash65}, where indeed the channel is shown on the cover of
the book.

The channel behaves as follows.  Balls labeled `$0$' or `$1$' are
used to communicate through the channel.  The channel starts with
a ball already in it. To use the channel, a ball is inserted into
the channel by the transmitter, and the receiver receives one of
the two balls in the channel with equal probability.  The ball
that does not exit the channel remains inside for the next channel
use.

Another appropriate name for this channel is {\it chemical
channel}\footnote{The name ``chemical channel'' is due to T.
Cover.}. This name suggests a physical system in which the
concentrations of chemicals are used to communicate, such as might
be the case in some cellular biological systems. The transmitter
adds molecules to the channel and the receiver samples molecules
randomly from the channel.  The trapdoor channel is the most basic
realization of this type of channel; it has only two types of
molecules and there are only three possible concentrations,
$(0,0.5,1)$, or alternatively only one molecule remains in the
channel between uses.

Although the trapdoor channel is very simple to describe, its
capacity has been an open problem for 45 years \cite{Blackwell61}.
The zero-error capacity was found by Ahlswede et al.
\cite{Ahl_kaspi87, Ahlswede98 } to be $0.5$ bits per channel use.
More recently, Kobayashi and Morita \cite{Kobayashi02} derived a
recursion for the conditional probabilities of output sequences of
length $n$ given the input sequences and used it to show that the
capacity of this channel is strictly larger than $0.5$ bits.
Ahlswede and Kaspi \cite{Ahl_kaspi87} considered two modes of the
channel called the {\it permuting jammer} channel and the {\it
permuting relay} channel. In the first mode there is a jammer in
the channel who attempts to frustrate the message sender by
selective release of balls in the channel. In the second mode,
where the sender is in the channel, there is a helper supplying
balls of a fixed sequence at the input, and the sender is
restricted to permuting this sequence. The helper collaborates
with the message sender in the channel to increase his ability to
transmit distinct messages to the receiver. Ahlswede and Kaspi
\cite{Ahl_kaspi87} gave answers for specific cases of both
situations and Kobayashi \cite{Kobayashi87} established the answer
to the general permuting relay channel. More results for specific
cases of the permuting jammer channel can be found in
\cite{piret89,Chan93}.

% The second sentence of this paragraph is confusing because it says "also."
In this paper we consider the trapdoor channel with feedback. We
derive the feedback capacity of the trapdoor channel by solving an
equivalent dynamic programming problem. Our work consists of two
main steps. The first step is formulating the feedback capacity of
the trapdoor channel as an infinite-horizon dynamic program,  and
the second step is finding explicitly the exact solution of that
program.

Formulating the feedback capacity problem as a dynamic program
appeared in Tatikonda's thesis \cite{Tatikonda00} and in work by
Yang, Kav\u{c}i\'{c} and Tatikonda \cite{Yang05}, Chen and Berger
\cite{Chen05}, and recently in a work by Tatikonda and Mitter
\cite{Tatikonda06}. Yang et. al.\cite{Yang05} have shown that if a
channel has a one-to-one mapping between the input and the state,
it is possible to formulate feedback capacity as a dynamic
programming problem and to find an approximate solution by using
the value iteration algorithm \cite{Bertsekas05}.  Chen and Berger
\cite{Chen05} showed that if the state of the channel is a
function of the output then it is possible to formulate the
feedback capacity as a dynamic program with a finite number of
states.

Our work provides the dynamic programming formulation and a
computational algorithm for finding the feedback capacity of a
family of channels called unifilar Finite State Channels (FSC's),
which include the channels considered in \cite{Yang05,Chen05}.
%Formulating the feedback
%capacity of a channel as a dynamic program allows us to use new
%methods to solve the problem.
We use value iteration \cite{Bertsekas05} to find an approximate
solution and to generate a conjecture for the exact solution, and
the Bellman equation \cite{Arapos93_average_cose_survey} to verify
the optimality of the conjectured solution.  As a result, we are
able to show that the feedback capacity of the trapdoor channel is
$\log\phi$, where $\phi$ is the golden ratio,
$\frac{1+\sqrt{5}}{2}$.  In addition, we present a simple
encoding/decoding scheme that achieves this capacity.
The remainder of the paper is organized as follows. Section
\ref{s_channel_model_and_pre} defines the channel setting and the
notation throughout the paper. Section \ref{s_main_results} states
the main results of the paper. Section
\ref{s_the_capacity_formula_for_unifilar_channel_with_feedback}
presents the capacity of a unifilar FSC in terms of directed
information. Section
\ref{s_feedback_capacity_and_dynamic_programming} introduces the
dynamic programming framework and shows that the feedback capacity
of the unifilar FSC can be characterized as the optimal average
reward of a dynamic program. Section
\ref{s_solution_for_the_chemical} shows an explicit solution for
the capacity of the trapdoor channel by using the dynamic
programming formulation. Section \ref{s_an_achievale_scheme} gives
a simple communication scheme that achieves the capacity of the
trapdoor channel with feedback and finally Section
\ref{s_conclusion} concludes this work.

%Section \ref{s_preliminaries} defines the channel setting and the
%notation throughout the paper. Section \ref{s_main_results} states
%the main results of the paper. Section
%\ref{s_random_generation_of_cod} defines and states several
%properties of the random generation of the coding scheme that is
%used in finding an achievable rate. Section \ref{s_acievable_rate}
%provides a bound on the error of the maximum likelihood decoder
%and uses this bound to find an achievable rate. Section
%\ref{s_upper_bound_on_capacity} gives an upper bound on the
%capacity and, finally, Section \ref{s_indeco} gives the capacity
%of an indecomposable FSC without intersymbol interference.

\section{Channel Models and Preliminaries} {\label{s_channel_model_and_pre} }We use subscripts and
superscripts to denote vectors in the following ways:
$x^j=(x_1\dots x_j)$ and $x_i^j=(x_i\dots x_j) $ for $i\leq j$.
Moreover, we use lower case $x$ to denote sample values, upper
case $X$ to denote random variables, calligraphic letter $\cal X$
to denote the alphabet and $|{\cal X}|$ to denote the cardinality
of the alphabet. The probability distributions are denoted by $p$
when the arguments specify the distribution, e.g.
$p(x|y)=p(X=x|Y=y)$. In this paper we consider only channels for
which the input, denoted by $\{X_1,X_2,...\}$, and the output,
denoted by $\{Y_1,Y_2,...\}$, are from finite alphabets, $\cal X$
and $\cal Y$, respectively. In addition, we consider only the
family of FSC known as unifilar channels as considered by
Ziv\cite{Ziv85}. An FSC is a channel that, for each time index,
has one of a finite number of possible states, $s_{t-1}$, and has
the property  that
$p(y_t,s_t|x^t,s^{t-1},y^{t-1})=p(y_t,s_t|x_t,s_{t-1})$. A {\it
unifilar} FSC also has the property that the state $s_t$ is
deterministic given $(s_{t-1},x_t,y_t)$:
\begin{definition}
An FSC is called a {\it unifilar FSC} if there exists a
time-invariant function $f(\cdot)$ such that the state evolves
according to the equation
\begin{equation}
s_t=f(s_{t-1},x_{t},y_{t}).
\end{equation}
\end{definition}
We also define a {\it strongly connected} FSC, as follows.
\begin{definition}
We say that a finite state channel is strongly connected if for
any state $s$ there exists an integer $T$ and an input
distribution of the form $\{p(x_t|s_{t-1})\}_{t=1}^T$ such that
the probability that the channel reaches state $s$ from any
starting state $s'$, in less than $T$ time-steps, is positive.
I.e.
\begin{equation}
\sum_{t=1}^T \Pr(S_t=s|S_0=s')>0,   \ \ \forall s\in {\cal
S},\forall s'\in {\cal S}.
\end{equation}
\end{definition}

\begin{figure}[h]{
\psfrag{v1\r}[][][0.8]{$m$}\psfrag{w1\r}[][][0.8]{Message}
 \psfrag{u1\r}[][][0.8]{Encoder}
\psfrag{d1\r}[][][0.8]{$x_t(m,y^{t-1})$}
\psfrag{v2\r}[][][0.8]{$x_t$} \psfrag{w2\r}[][][0.8]{$$}
 \psfrag{u2\r}[][][0.8]{Unifilar Finite State Channel}
\psfrag{d2a\r}[][][0.8]{$p(y_t|x_t,s_{t-1})$}
\psfrag{d2b\r}[][][0.8]{$s_t=f(s_{t-1},x_{t},y_{t})$}
\psfrag{v3\r}[][][0.8]{$y_t$}
\psfrag{w3\r}[][][0.8]{$$}\psfrag{v1\r}[][][0.8]{$m$}
\psfrag{u3\r}[][][0.8]{Decoder} \psfrag{d3\r}[][][0.8]{$\hat
m(y^N)$} \psfrag{u5\r}[][][1]{} \psfrag{d5\r}[][][0.8]{$y_{t}$}
\psfrag{v4\r}[][][0.8]{$y_t$} \psfrag{w4\r}[][][0.8]{$$}
\psfrag{u4\r}[][][0.8]{Unit Delay}
\psfrag{v5\r}[][][0.8]{$y_{t-1}$}
 \psfrag{w5\r}[][][0.8]{Feedback}
\psfrag{w6b\r}[][][0.8]{message} \psfrag{v6\r}[][][0.8]{$\hat
m$}\psfrag{w6a\r}[][][0.8]{Estimated}
\centerline{\includegraphics[width=12cm]{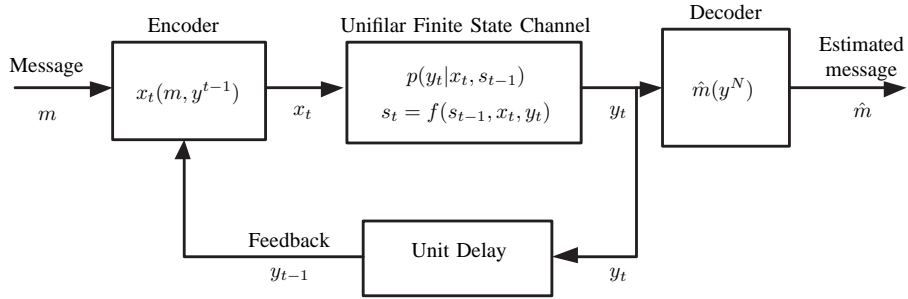}}
\caption{Unifilar FSC with feedback} \label{f_1}
}\end{figure} %%when I save on visio I used AI 3 - and I had to restart the computer and not all visio features work with AI3

We assume a communication setting that includes feedback as shown
in Fig. \ref{f_1}. The transmitter (encoder) knows at time $t$ the
message $m$ and the feedback samples $y^{t-1}$. The output of the
encoder at time $t$ is denoted by $x_t$ and is a function of the
message and the feedback. The channel is a unifilar FSC and the
output of the channel $y_t$ enters the decoder (receiver). The
encoder receives the feedback sample with one unit delay.

\subsection{Trapdoor Channel is a Unifilar FSC}
%The chemical channel, which is shown in figure
%\ref{f_chemical_horizontal}, is also called the trap-door channel
%and was introduced in 1961 by Blackwell \cite{Blackwell61} as a
%simple two state channel. A detailed discussion of this channel
%appears
% in a book by Ash \cite{Ash65}.
%The chemical channel acts as follows: The channel starts with one
%ball labeled 0 or 1, and after receiving another ball labeled also
%with a number 0 or 1 the channel outputs one of the balls with
%equal probability.
The state of the trapdoor channel, which is described in the
introduction and shown in figure \ref{f_chemical_horizontal}, is
the ball, 0 or 1, that is in the channel before the transmitter
transmits a new ball. Let $x_t\in\{0,1\}$ be the ball that is
transmitted at time $t$ and $s_{t-1}\in\{0,1\}$ be the state of
the channel when ball $x_t$ is transmitted. The probability of the
output $y_t$ given the input $x_t$ and the state of the channel
$s_{t-1}$ is shown in table \ref{ta_prob_chem}.

\begin{table}[h]
\caption{The probability of the output $y_t$ given the input $x_t$
and the state $s_{t-1}$.} \centering \label{ta_prob_chem}
\begin{tabular}[h]{||c|c|c|c||}
\hline \hline
$x_t$ & $s_{t-1}$ & $p(y_t=0|x_t,s_{t-1})$ & $p(y_t=1|x_t,s_{t-1})$\\
\hline \hline
0 & 0 & 1 & 0 \\
\hline
0 & 1 & 0.5 & $0.5$ \\
\hline
1 & 0 & $0.5$ & $0.5$\\
\hline
1 & 1 & 0 & 1 \\
\hline \hline
\end{tabular}
\end{table}
The trapdoor channel is a unifilar FSC. It has the property that
the next state $s_t$ is a deterministic function of the state
$s_{t-1}$, the input $x_t$, and the output $y_t$. For a feasible
tuple, $(x_t,y_t,s_{t-1})$, the next state is given by the
equation
\begin{equation}\label{e_chemical_st}
s_t=s_{t-1}\oplus x_t \oplus y_t,
\end{equation}
where $\oplus$ denotes the binary XOR operation.

\subsection{Trapdoor Channel is a Permuting Channel}
\begin{figure}[h]{
\centerline{\includegraphics[width=7cm]{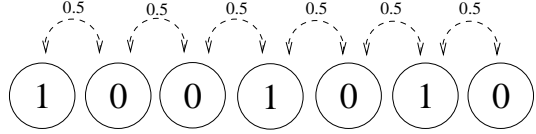}}
%\centerline{\includegraphics[width=7cm]{C:/cygwin/usr/X11R6/bin/home/chemical_permuting.eps}}
\caption{The trapdoor channel as a permuting channel. Going from
left to right, there is a probability of one half that two
adjacent bits switch places.} \label{f_chemical_permuting} }
\end{figure}

It is interesting to note, although not consequential in this
paper, that  the trapdoor channel is a permuting channel
\cite{Benjamin75}, where the output is a permutation of the input
(Fig. \ref{f_chemical_permuting}). At each time $t$, a new bit is
added to the sequence and the channel switches the new bit with
the previous one in the sequence with probability 0.5.
%I should write an email to Toby berger about it

\section{Main Results}\label{s_main_results}
\begin{itemize}
\item
The capacity of the trapdoor channel with feedback is
\begin{equation}
C=\log \frac{\sqrt 5 +1}{2}.
\end{equation}
Furthermore, there exists a simple capacity achieving scheme which
will be presented in Section \ref{s_an_achievale_scheme}.

\item
The problem of finding the capacity of a strongly connected
unifilar channel (Fig. \ref{f_1}) can be formulated as an
average-reward dynamic program, where the state of the dynamic
program is the probability mass function over the states
conditioned on prior outputs, and the action is the stochastic
matrix $p(x|s)$. By finding a solution to the average-reward
Bellman equation we find the exact capacity of the channel.

\item
As a byproduct of our analysis we also derive a closed form
solution to an infinite horizon average-reward dynamic program
with a continuous state-space.
%\subsection{Example: additive noise with memory}
\end{itemize}

\section{The Capacity Formula for a Unifilar Channel with
Feedback}\label{s_the_capacity_formula_for_unifilar_channel_with_feedback}

The main goal of this section is to prove the following theorem
which allows us to formulate the problem as a dynamic program.
\begin{theorem}\label{t_feedback_capacity_unifilar_FSC1}
The feedback capacity of a strongly connected unifilar FSC when
initial state $s_0$ is known at the encoder and decoder can be
expressed as

\begin{equation}\;
C_{FB}=
%\max_{{\{p(x_{t}|s_{t-1},y^{t-1})\}_{t=1}^N}}\sum_{t=1}^{N}
%I(X_t,S_{t-1};Y_t|Y^{t-1}),
\sup_{\{p(x_{t}|s_{t-1},y^{t-1})\}_{t\geq 1}} \liminf_{N
\rightarrow \infty} \frac{1}{N}\sum_{t=1}^{N}
I(X_t,S_{t-1};Y_t|Y^{t-1})
\end{equation}
 where $\{p(x_{t}|s_{t-1},y^{t-1})\}_{t\geq1}$ denotes the set
of all distributions such that
$p(x_t|y^{t-1},x^{t-1},s^{t-1})=p(x_t|s_{t-1},y^{t-1})$ for
$t=1,2,...$ .
\end{theorem}

Theorem \ref{t_feedback_capacity_unifilar_FSC1} is a direct
consequence of Theorem \ref{t_feedback_capacity_unifilar_FSC} and
eq. (\ref{e_lequality2}) in Lemma \ref{l_FSC_equality}, which are
proved in this section.
%\begin{proof}
%The proof is straightforward by applying equality
%(\ref{e_lequality}) from Lemma \ref{l_FSC_equality} to Theorem
%\ref{t_feedback_capacity_unifilar_FSC}. Both Lemma
%\ref{l_FSC_equality} and Theorem
%\ref{t_feedback_capacity_unifilar_FSC} are proven in this section.
%\end{proof}

%This capacity formulation allows us to formulate the problem as a
%dynamic programming as shown in the following section.

For any finite state channel with perfect feedback, as shown in
Figure \ref{f_1}, the capacity was shown in
\cite{PermuterISIT06,Permuter06_feedback_submit} to be bounded as
\begin{equation} \label{e_capacity_feedback_sand}
 \lim_{N \rightarrow \infty}
\frac{1}{N}  \max_{p(x^N||y^{N-1})} \max_{s_0}I(X^N \rightarrow
Y^N|s_0 ) \geq C_{FB} \geq \lim_{N \rightarrow \infty} \frac{1}{N}
\max_{p(x^N||y^{N-1})} \min_{s_0}I(X^N \rightarrow Y^N |s_0).
\end{equation}
The term $I(X^N \rightarrow Y^N )$ is the {\it directed
information}\footnote{In addition to feedback capacity, directed
information has recently been used in rate distortion
\cite{Pradhan04}, \cite{Pradhan04b}, \cite{zamir06}, network
capacity \cite{Kramer03}, \cite{Kramer88}  and computational
biology \cite{Hero06}.} defined originally by Massey in
\cite{Massey90} as
\begin{equation} I(X^N \rightarrow Y^N )\triangleq
\sum_{t=1}^{N}I(X^t;Y_t|Y^{t-1}).
\end{equation}
The initial state is denoted as $s_0$ and $p(x^N||y^{N-1})$ is the
causal conditioning distribution defined
\cite{Kramer03,PermuterISIT06} as
\begin{equation}
p(x^N||y^{N-1})\triangleq \prod_{t=1}^{N} p(x_t|x^{t-1},y^{t-1}).
\end{equation}
The directed information in eq. (\ref{e_capacity_feedback_sand})
is under the distribution of $p(x^n,y^n)$ which is uniquely
determined by the causal conditioning, $p(x^N||y^{N-1})$, and by
the channel.

In our communication setting we are assuming that the initial
state is known both to the decoder and to the encoder. This
assumption allows the encoder to know the state of the channel at
any time $t$ because $s_t$ is a deterministic function of the
previous state, input and output. In order to take into account
this assumption, we use a trick of allowing a fictitious time
epoch before the first actual use of the channel in which the
input does not influence the output nor the state of channel and
the only thing that happens is that the output equals $s_0$ and is
fed back to the encoder such that at time $t=1$ both the encoder
and the decoder know the state $s_0$. Let $t=0$ be the fictitious
time before starting the use of the channel. According to the
trick, $Y_{0}=S_0$ and the input $X_{0}$ can be chosen arbitrarily
because it does not have any influence whatsoever. For this
scenario the directed information term in eq.
(\ref{e_capacity_feedback_sand}) becomes
\begin{equation}
I(X_0^N \rightarrow Y_0^N|s_0)=I(X^N \rightarrow Y^N |s_0).
\end{equation}
%and
%\begin{equation}
%I(X_0^N \rightarrow Y_0^N)=I(X^N \rightarrow Y^N |S_0).
%\end{equation}
The input distribution becomes
\begin{equation}
 p(x_0^N||\{s_0, y^{N-1}\})= p( x^N||y^{N-1},s_0),
\end{equation}
where $p( x^N||y^{N-1},s_0)$ is defined as  $p( x^N||y^{N-1},s_0)
\triangleq \prod_{t=1}^N p(x_t|x^{t-1},y^{t-1},s_0)$. Therefore,
the capacity of a channel with feedback for which the initial
state, $s_0$, is known both at the encoder and the decoder is
bounded as
\begin{equation}\label{e_capacity_feedback_sandwitch_s0}
 \lim_{N \rightarrow \infty}
\frac{1}{N}  \max_{p(x^N||y^{N-1},s_0)} \max_{s_0}I(X^N
\rightarrow Y^N |s_0)  \geq C_{FB} \geq \lim_{N \rightarrow
\infty} \frac{1}{N} \max_{p(x^N||y^{N-1},s_0)} \min_{s_0}I(X^N
\rightarrow Y^N |s_0)
\end{equation}

\begin{lemma}\label{l_strongly_connected_s0}
If the finite state channel is strongly connected, then for any
input distribution $p_1(x^N||y^{N-1},s_0)$ and any $s_0'$ there
exists an input distribution $p_2(x^N||y^{N-1},s'_0)$ such that
\begin{equation}
 \frac{1}{N} \left|  I_{p_1}(X^N
\rightarrow Y^N |s_0) -  I_{p_2}(X^N \rightarrow Y^N
|s'_0)\right|\leq \frac{c}{N}
\end{equation}
where $c$ is a constant that does not depend on $N$, $s_0$,
$s_0'$. The term $I_{p_1}(X^N \rightarrow Y^N |s_0)$ denotes the
directed information induced by the input distribution
$p_1(x^N||y^{N-1},s_0)$ where $s_0$ is the initial state.
Similarly, the term $I_{p_2}(X^N \rightarrow Y^N |s_0')$ denotes
the directed information induced by the input distribution
$p_2(x^N||y^{N-1},s_0')$ where $s_0'$ is the initial state.
\end{lemma}

\begin{proof}
Construct ${p_2}(x^N||y^N,s_0')$ as follows. Use an input
distribution, which has a positive probability of reaching $s_0$
in $T$ time epochs, until the time that the channel first reaches
$s_0$. Such an input distribution exists because the channel is
strongly connected. Denote the first time that the state of the
channel equals $s_0$ by $L$. After time $L$, operate exactly as
$p_1$ would (had time started then). Namely, for $t>L$,
$p_2(x_{t}|x^{t-1},y^{t-1},s_0)=p_1(x_{t-L}|x^{t-L-1},y^{t-L-1},s_0)$.
Then
% Let $L$ be the time
%that the channel first reaches $s_0$. For the first $L$ time
%epochs, ${p_2}(x^L||y^L,s_0')$ will be a policy that has a
%positive probability of reaching the state $s_0$. Namely, for
%$t\leq L$, $p_2(x_t|x^{t-1},y^{t-1},s_0')=p(x_t|s_{t-1})$ where
%$p(x_t|s_{t-1})$ is a distribution that reaches $s_0$ with
%positive probability in a bounded time. Then, beginning at time
%$L$, ${p_2}(x_L^{N+L}||y_L^{N+L} ,s_0')$ will be identical to
%${p_1}(x^N ||y^N |s_0')$, that is for $t>L$,
%$p_2(x_{t}|x^{t-1},y^{t-1},s_0)=p_1(x_{t-L}|x^{t-L-1},y^{t-L-1},s_0)$.
%For the first $L$ epoch times ${p_2}(X^L||Y^L |s_0')$ will be a
%policy that has a positive probability to reach the state $s_0$
%where $L$ is the time that the channel first reached $s_0$. Then
%${p_2}(X_L^{N+L}||Y_L^{N+L} |s_0')$ will be identical to
%${p_1}(X^N ||Y^N |s_0')$. Namely, for $t\leq L$,
%$p_2(x_t|x^{t-1},y^{t-1},s_0')=p(x_t|s_{t-1})$ where
%$p(x_t|s_{t-1})$ is a distribution that reaches $s_0$ with
%positive probability in a bounded time and for $t>L$,
%$p_2(x_{t}|x^{t-1},y^{t-1},s_0)=p_1(x_{t-L}|x^{t-L-1},y^{t-L-1},s_0)$.
\begin{eqnarray}
\lefteqn{ \frac{1}{N}\left|I_{p_1}(X^N \rightarrow Y^N |s_0) -
I_{p_2}(X^N \rightarrow Y^N |s'_0)\right|}
\nonumber \\
&\stackrel{(a)}{\leq}& \frac{1}{N}\left| I_{p_1}(X^N \rightarrow
Y^N |s_0) -  I_{p_2}(X^N \rightarrow Y^N |L,s'_0)\right|+
\frac{1}{N} H(L)
\nonumber \\
&\stackrel{(b)}{=}& \frac{1}{N}\left|
\sum_{l=1}^{\infty}p(L=l)I_{p_1}(X^N \rightarrow Y^N |s_0)-
\sum_{l=1}^{\infty}p(L=l)\left(I_{p_2}(X^{N}_l \rightarrow Y^{N}_l
|s_l)+  I_{p_2}(X^l \rightarrow Y^l |s_l,s'_0)\right) \right|+
\frac{1}{N} H(L)
\nonumber \\
&\stackrel{(c)}{\leq}& \frac{1}{N}\left|
\sum_{l=1}^{\infty}p(L=l)I_{p_1}(X^N \rightarrow Y^N |s_0)-
\sum_{l=1}^{\infty}p(L=l) I_{p_2}(X^{N}_l \rightarrow Y^{N}_l
|s_l)\right|\nonumber \\
&&+ \frac{1}{N}
 \left|\sum_{l=1}^{\infty}p(L=l) I_{p_2}(X^l \rightarrow Y^l |s_l,s'_0)
\right|+ \frac{1}{N} H(L)
\nonumber \\
%&\stackrel{(d)}{\leq}& \frac{1}{N}
% \left|\sum_{l=1}^{\infty}p(L=l) I_{p_1}(X_{N-l+1}^N \rightarrow Y^l |s_0)
%\right|+ \frac{1}{N}
% \left|\sum_{l=1}^{\infty}p(L=l) I_{p_2}(X^l \rightarrow Y^l |s_l,s'_0)
%\right|+ \frac{1}{N} H(L)
%\nonumber \\
&\stackrel{(d)}{\leq}& \frac{2}{N}\sum_{l=1}^{\infty}p(L=l) l
\log |\mathcal{Y}|+ \frac{1}{N} H(L)
\nonumber \\
&\stackrel{}{=}& \frac{1}{N} \left( \log |\mathcal{Y}|
\mathbb{E}[L] + H(L) \right)
%&\stackrel{}{\leq}& \frac{1}{N}\left( \log| \mathcal{Y}|
%\frac{1}{p}+  H(L)\right)
\end{eqnarray}

\begin{itemize}
\item[(a)] from the triangle inequality and Lemma 3 in
\cite{PermuterISIT06} which claims that for an arbitrary random
variables $(X^N,Y^N,S)$, the inequality $ \left|I(X^N \rightarrow
Y^N)- I(X^N \rightarrow Y^N|S)\right|\leq H(S)\ $ always holds.

\item[(b)] follows from  using the special structure of
$p_2(x^N||y^N,s_0')$.

\item[(c)] follows from the triangle inequality.

\item[(d)] follows from the fact that in the first absolute
value $N-l$ terms cancel and therefor only $l$ terms remain where
each one of them is bounded by $I(X^t;Y_t|Y^{t-1})\leq
|\mathcal{Y}|$. In the second absolute value there are $l$ terms
also bounded by $|\mathcal{Y}|$.

\end{itemize}
The proof is completed by noting that $H(L)$ and $E(L)$ are upper
bounded respectively, by $H(\tilde L)$ and $E(\tilde L)$ where $
\lfloor \tilde L/T \rfloor \sim \text{Geometric}(p)$ and $p$ is
the minimum probability of reaching $s_0$ in less than $T$ steps
from any state $s\in {\cal S}$. Because the random variable
$\lfloor\tilde L/T\rfloor$ has a geometric distribution, $H(\tilde
L)$ and $\mathbb{E}[\tilde L]$ are finite and, consequently, so
are $H(L)$ and $E(L)$.
\end{proof}

\begin{theorem}\label{t_feedback_capacity_unifilar_FSC}
The feedback capacity of a strongly connected unifilar FSC when
initial state is known at the encoder and decoder is given by

\begin{equation}
C_{FB}=\lim_{N \rightarrow \infty} \frac{1}{N}
\max_{{\{p(x_{t}|s_{t-1},y^{t-1})\}_{t=1}^N}}\sum_{t=1}^{N}
I(X_t,S_{t-1};Y_t|Y^{t-1}),
\end{equation}
\end{theorem}

\begin{proof}
The proof of the theorem contains four main equalities which are
proven separately.
\begin{eqnarray}
C_{FB}&\stackrel{}{=}&\lim_{N \rightarrow \infty} \frac{1}{N}
\max_{p(x^N||y^{N-1},s_0)} \min_{s_0} I(X^N \rightarrow Y^N
|s_0) \label{e_theorem_step0}\\
&\stackrel{}{=}&\lim_{N \rightarrow \infty} \frac{1}{N}
\max_{p(x^N||y^{N-1},s_0)}I(X^N \rightarrow Y^N |S_0) \label{e_theorem_step1}\\
&\stackrel{}{=}&\lim_{N \rightarrow \infty} \frac{1}{N}
\max_{p(x^N||y^{N-1},s_0)}\sum_{t=1}^{N}
I(X_t,S_{t-1};Y_t|Y^{t-1})
\label{e_theorem_step2} \\
&\stackrel{}{=}&\lim_{N \rightarrow \infty} \frac{1}{N}
\max_{{\{p(x_{t}|s_{t-1},y^{t-1})\}_{t=1}^N}}\sum_{t=1}^{N}
I(X_t,S_{t-1};Y_t|Y^{t-1}). \label{e_theorem_step3}
\end{eqnarray}

{\it Proof of equality (\ref{e_theorem_step0}) and
(\ref{e_theorem_step1}):} As a result of Lemma
\ref{l_strongly_connected_s0},

\begin{eqnarray}
\lim_{N \rightarrow \infty} \frac{1}{N} \max_{p(x^N||y^{N-1},s_0)}
I(X^N \rightarrow Y^N |S_0)&\stackrel{(a)}{=}& \lim_{N \rightarrow
\infty} \frac{1}{N} \max_{p(x^N||y^{N-1},s_0)} \sum_{s_0}p(s_0)
I(X^N \rightarrow Y^N|s_0)
\label{e_mins0a} \nonumber \\
&\stackrel{(b)}{=}& \lim_{N \rightarrow \infty} \frac{1}{N}
\sum_{s_0}p(s_0)\max_{p(x^N||y^{N-1},s_0)}  I(X^N \rightarrow
Y^N|s_0)
\label{e_mins0b}\nonumber \\
&\stackrel{(c)}{=}& \lim_{N \rightarrow \infty} \frac{1}{N}
\min_{s_0} \max_{p(x^N||y^{N-1},s_0)}  I(X^N \rightarrow Y^N|s_0)
 \label{e_mins0c}\\
&\stackrel{(d)}{=}& \lim_{N \rightarrow \infty} \frac{1}{N}
\max_{p(x^N||y^{N-1},s_0)} \min_{s_0} I(X^N \rightarrow Y^N
|s_0).\label{e_mins0}
\end{eqnarray}
where,
\begin{itemize}
\item[(a)] follows from the definition of conditional entropy.
\item[(b)] follows from exchanging between the summation and the maximization. The exchange is possible because maximization is over causal conditioning
distributions that depend on $s_0$ .
\item[(c)]
follows from Lemma \ref{l_strongly_connected_s0}.
\item[(d)] follows from the observation that the distribution $p^*(x^N||y^{N-1},s_0)$ that achieves the maximum in (\ref{e_mins0c}) and in
(\ref{e_mins0}) is the same: $p^*(x^N||y^{N-1},s_0)=\arg
 \max_{p(x^N||y^{N-1},s_0)}  I(X^N \rightarrow Y^N|s_0)$. This
 observation allows us to exchange the order of the minimum and the maximum.
\end{itemize}
Equations (\ref{e_mins0c}) and (\ref{e_mins0}) can be repeated
also with $\max_{s_0}$ instead of $\min_{s_0}$ and hence we get
\begin{eqnarray}
\lim_{N \rightarrow \infty} \frac{1}{N} \max_{p(x^N||y^{N-1},s_0)}
I(X^N \rightarrow Y^N |S_0)&\stackrel{}{=}& \lim_{N \rightarrow
\infty} \frac{1}{N} \max_{p(x^N||y^{N-1},s_0)} \max_{s_0} I(X^N
\rightarrow Y^N |s_0).\label{e_maxs0}
\end{eqnarray}

By using eq. (\ref{e_mins0}) and (\ref{e_maxs0}), we get that the
upper bound and lower bound in
(\ref{e_capacity_feedback_sandwitch_s0}) are equal and therefore
eq. (\ref{e_theorem_step0}) and (\ref{e_theorem_step1}) hold.
%\begin{eqnarray}\label{e_c_fb}
%C_{FB}&\stackrel{}{=}& \lim_{N \rightarrow \infty} \frac{1}{N}
%\max_{p(x^N||y^{N-1},s_0)} \min_{s_0} I(X^N \rightarrow Y^N
%|s_0)\nonumber \\
% &=&\lim_{N \rightarrow \infty} \frac{1}{N}
%\max_{p(x^N||y^{N-1},s_0)}I(X^N \rightarrow Y^N |S_0).
%\end{eqnarray}

{\it Proof of equality (\ref{e_theorem_step2}):} Using the
property that the next state of the channel is a deterministic
function of the input, output and current state we get,
\begin{eqnarray}\label{e_cfb_simpler}
I(X^N \rightarrow Y^N |S_0) &=& \sum_{t=1}^{N}
I(X^t;Y_t|Y^{t-1},S_0)
\nonumber \\
&\stackrel{}{=}&\sum_{t=1}^{N}
H(Y_t|Y^{t-1},S_0)-H(Y_t|X^t,Y^{t-1},S_0)
\nonumber \\
&\stackrel{(a)}{=}&\sum_{t=1}^{N}
H(Y_t|Y^{t-1},S_0)-H(Y_t|X^t,Y^{t-1},S_0,S^{t-1}(X^t,Y^{t-1},S_0))
\nonumber \\
&\stackrel{(b)}{=}&\sum_{t=1}^{N}
H(Y_t|Y^{t-1},S_0)-H(Y_t|X_t,S_{t-1},Y^{t-1},S_0)
\nonumber \\
&\stackrel{}{=}&\sum_{t=1}^{N} I(S_{t-1},X_t;Y_t|Y^{t-1},S_0).
\end{eqnarray}
Equality (a) is due to the fact that $s^{t-1}$ is a deterministic
function of the tuple $(x^t,y^{t-1},s_0)$. Equality (b) is due to
the fact that
$p(y_t|x^t,y^{t-1},s^{t-1},s_0)=p(y_t|x_t,y^{t-1},s_{t-1},s_0)$.
By combining eq. (\ref{e_theorem_step1}) and eq.
(\ref{e_cfb_simpler}) we get eq. (\ref{e_theorem_step2}).
%\end{proof}

%The distribution that influence the
%directed information is $P(s_{t-1},x_t,y^{t})$ .
%\begin{theorem}\label{t_input_dist_sxy_channel}
%For a channel with finite memory as described in this section it
%is enough to consider an input distribution of the form
%$P(x_{t}|s_{t-1},y^{t-1})$ in order to achieve the capacity of the
%channel.
%\end{theorem}
%
%\begin{proof}
{\it Proof of equality (\ref{e_theorem_step3}):} It will suffice
to prove by induction that if we have two input distributions
$\{p_1(x_t|x^{t-1},y^{t-1},s_0)\}_{t\geq1}$ and
$\{p_2(x_t|x^{t-1},y^{t-1},s_0)\}_{t\geq1}$ that induce the same
distributions $\{p(x_t|s_{t-1},y^{t-1})\}_{t\geq1}$ then the
distributions $\{p(s_{t-1},x_t,y^{t})\}_{t\geq1}$ are equal under
both inputs. First let us verify the equality for $t=1$:
\begin{eqnarray}
p(s_{0},x_1,y_{1})=p(s_0)p(x_1|s_0)p(y_1|s_0,x_1).
\end{eqnarray}
Since $p(s_0)$ and $p(y_1|s_0,x_1)$ are not influenced by the
input distribution and since $p(x_1|s_0)$ is equal for both input
distributions then $p(s_{0},x_1,y_{1})$ is also for both input
distributions. Now, we assume that $p(s_{t-1},x_t,y^{t})$ is equal
under both input distributions and we need to prove that
$p(s_{t},x_{t+1},y^{t+1})$ is also equal under both input
distributions. The term $p(s_{t},x_{t+1},y^{t+1})$ which can be
written as,
\begin{eqnarray}
p(s_{t},x_{t+1},y^{t+1}) & = &
p(s_t,y^t)p(x_{t+1}|s_t,y^t)p(y_{t+1}|x_{t+1},s_t).
\end{eqnarray}
First we notice that if $p(s_{t-1},x_t,y^{t})$ is equal for both
cases then necessarily $p(s_{t-1},s_t,x_t,y^{t})$ is also equal
for both cases because $s_t$ is a deterministic function of the
tuple $(s_{t-1},x_t,y_{t})$ and therefore both input distributions
induce the same $p(s_t,y^t)$. The distribution,
$p(x_{t+1}|s_t,y^t)$, is the same under both input distributions
by assumption and $p(y_{t+1}|x_{t+1},s_t)$ does not depend on the
input distribution.
\end{proof}

The next lemma shows that it is possible to switch between the
limit and the maximization in the capacity formula. This is
necessary for formulating the problem, as we do in the next
section, as an average-reward dynamic program.

\begin{lemma}\label{l_FSC_equality}
For any FSC the following equality holds:
\begin{equation}\label{e_lequality}
\lim_{N \rightarrow \infty} \frac{1}{N}
\max_{p(x^N||y^{N-1},s_0)}\min_{s_0}I(X^N \rightarrow Y^N
|s_0)=\sup_{\{P(x_t|y^{t-1},x^{t-1},s_0)\}_{t\geq 1}} \liminf_{N
\rightarrow \infty} \frac{1}{N} \min_{s_0} I(X^N \rightarrow Y^N
|s_0).
\end{equation}
And, in particular, for a strongly connected unifilar FSC
\begin{equation}\label{e_lequality2}
\lim_{N \rightarrow \infty} \frac{1}{N}
\max_{{\{p(x_{t}|s_{t-1},y^{t-1})\}_{t=1}^N}}\sum_{t=1}^{N}
I(X_t,S_{t-1};Y_t|Y^{t-1})
%\lim_{N \rightarrow \infty}
%\frac{1}{N}
%\max_{p(x^N||y^{N-1},s_0)}\min_{s_0}I(X^N \rightarrow Y^N |s_0)
=\sup_{\{p(x_{t}|s_{t-1},y^{t-1})\}_{t\geq 1}} \liminf_{N
\rightarrow \infty} \frac{1}{N}\sum_{t=1}^{N}
I(X_t,S_{t-1};Y_t|Y^{t-1})
\end{equation}
%where the supremium is over the set of distributions,
%$\{p(x^t|y^{t-1},x^{t-1},s_0)\}_{t\geq1}$.
\end{lemma}
 On the left-hand side of the
equations appears $\lim$ because, as shown in
\cite{Permuter06_feedback_submit}, the limit exists due to the
super-additivity property of the sequence.

\begin{proof}
We are going to prove eq. (\ref{e_lequality}) which hold for any
FSC. For the case of unifilar channel, the left-hand side of eq.
(\ref{e_lequality})
 is proven to be equal to the left side of eq.
 (\ref{e_lequality2}) in eq.
(\ref{e_theorem_step0})-(\ref{e_theorem_step3}). By the same
arguments as in (\ref{e_theorem_step0})-(\ref{e_theorem_step3})
also the right-hand side of (\ref{e_lequality}) and
(\ref{e_lequality2}) are equal.

  Define
 \begin{equation}
 \underline C_N \triangleq \frac{1}{N}  \max_{p(x^N||y^{N-1},s_0)} \min_{s_0}I(X^N
\rightarrow Y^N |s_0).
 \end{equation}

In order to prove that the equality holds we will use two
properties of $\underline C_N$ that were proved in \cite[Theorem
13]{Permuter06_feedback_submit}.

The first property, is that $ \underline C_N$ is a super additive
sequence, namely,
\begin{eqnarray}
N \left[\underline C_N -\frac{\log |\mathcal{S}|}{N}\right] \geq
n\left[\underline C_n - \frac{\log |\mathcal{S}|}{n}\right] +
l\left[\underline C_l-\frac{\log |\mathcal{S}|}{l}\right].
\end{eqnarray}

%\begin{eqnarray}
%N \underline C_N & \geq & n\underline C_n +  l\underline C_l-\log
%|{\cal S}|.
%\end{eqnarray}
The second property, which is a result of the first, is that
 \begin{equation}
  \lim_{N \rightarrow \infty} \underline C_N= \sup_{N} \underline C_N
 \end{equation}

Now, consider
\begin{eqnarray}
\lim_{N \rightarrow \infty} \frac{1}{N}
\max_{p(x^N||y^{N-1},s_0)}\min_{s_0}I(X^N \rightarrow Y^N
|s_0)&\stackrel{}{=}&\sup_{N}\underline C_N
\nonumber \\
&\stackrel{}{=}&\sup_{N} \frac{1}{N}
\max_{p(x^N||y^{N-1},s_0)}\min_{s_0}I(X^N \rightarrow Y^N |s_0)
\nonumber \\
&\stackrel{}{=}&\sup_{N} \frac{1}{N}
\sup_{\{p(x^t|y^{t-1},x^{t-1},s_0)\}_{t \geq 1}}\min_{s_0}I(X^N
\rightarrow Y^N |s_0)
\nonumber \\
&\stackrel{}{=}& \sup_{\{p(x^t|y^{t-1},x^{t-1},s_0)\}_{t \geq 1}}
\sup_{N} \frac{1}{N} \min_{s_0}I(X^N \rightarrow Y^N |s_0)
\nonumber \\
&\stackrel{}{\geq}& \sup_{\{p(x^t|y^{t-1},x^{t-1},s_0)\}_{t \geq
1}} \liminf_{N} \frac{1}{N} \min_{s_0}I(X^N \rightarrow Y^N |s_0)
\end{eqnarray}

The limit of the left side of the equation in the lemma implies
that, $\forall \epsilon>0$ there exists $N(\epsilon)$ such that
for all $n>N(\epsilon)$,
$\frac{1}{n}\max_{p(x^n||y^{n-1},s_0)}\min_{s_0}I(X^N \rightarrow
Y^N |s_0)\geq \sup_N C_N-\epsilon$. Let us choose $j>N(\epsilon)$
and let $p^*(x^j||y^{j-1})$ be the input distribution that attains
the maximum. Let us construct

\begin{equation}
\tilde
p(x^t||y^{t-1},s_0)=p^*(x^t_{t-j+1}||y^{t-1}_{t-j+1},s_{t-j})p^*(x^{t-j}_{t-2j+1}||y^{t-j-1}_{t-2j+1},s_{t-2j})...\qquad.
\end{equation}

Then we get,

\begin{eqnarray}
\sup_{\{p(x^t|y^{t-1},x^{t-1},s_0)\}} \liminf_{N \rightarrow
\infty} \frac{1}{N} \min_{s_0} I(X^N \rightarrow Y^N
|s_0)\geq\liminf_{N \rightarrow \infty} \frac{1}{N} \min_{s_0}
I_{\tilde p}(X^N \rightarrow Y^N |s_0)\geq \sup_N C_N-\epsilon
\end{eqnarray}
where $I_{\tilde p}(X^N \rightarrow Y^N |s_0)$ is the directed
information induced by the input  $\tilde p(x^t||y^{t-1},s_0)$ and
the channel. The left inequality holds because $\tilde
p(x^t||y^{t-1},s_0)$ is only one possible input distribution among
all $\{p(x^t||y^{t-1},s_0)\}_{t=1}^{\infty}$. The right inequality
holds because the special structure of $\tilde
p(x^t||y^{t-1},s_0)$ transforms the whole expression of normalized
directed information into an average of infinite sums of terms
that each term is directed information between blocks of length
$j$. Because for each block the inequality holds, then it holds
also for the average of the blocks. The inequality may not hold on
the last block, but because we average over an increasing number
of blocks its influence diminishes.

\end{proof}

\section{Feedback Capacity and Dynamic
Programming}\label{s_feedback_capacity_and_dynamic_programming}

In this section, we characterize the feedback capacity of the
unifilar FSC as the optimal average-reward of a dynamic program.
Further, we present the Bellman equation, which can be solved to
determine this optimal average reward.

\subsection{Dynamic Programs}

Here we introduce a formulation for average-reward dynamic
programs. Each problem instance is defined by a septuple $({\cal
Z}, {\cal U}, {\cal W}, F, P_z, P_w, g)$.  We will explain the
roles of these parameters.

We consider a discrete-time dynamic system evolving according to
\begin{equation} \label{e_dp_state_evolution}
z_t = F(z_{t-1},u_t,w_t), \qquad t=1,2,3,\ldots,
\end{equation}
where each {\it state} $z_t$ takes values in a Borel space
${\mathcal Z}$, each {\it action} $u_t$ takes values in a compact
subset ${\cal U}$ of a Borel space, and each disturbance $w_t$
takes values in a measurable space ${\mathcal W}$. The initial
state $z_0$ is drawn from a distribution $P_z$.  Each disturbance
$w_t$ is drawn from a distribution $P_w(\cdot | z_{t-1}, u_t)$
which depends only on the state $z_{t-1}$ and action $u_t$.
%Note that, as opposed to $p(\cdot)$ which denotes probability with
%respect to an underlying probability space, $P_z$ and $P_w$ should
%be viewed as functions.
All functions considered in this paper
are assumed to be measurable, though we will not mention this each
time we introduce a function or set of functions.

The  {\it history} $h_t = (z_0, w_0, \ldots, w_{t-1})$ summarizes
information available prior to selection of the $t$th action.  The
action $u_t$ is selected by a function $\mu_t$ which maps
histories to actions.  In particular, given a policy $\pi =
\{\mu_1, \mu_2, \ldots\}$, actions are generated according to $u_t
= \mu_t(h_t)$.  Note that given the history $h_t$ and a {\it
policy} $\pi = \{\mu_1, \mu_2, \ldots\}$, one can compute past
states $z_1, \ldots, z_{t-1}$ and actions $u_1, \ldots, u_{t-1}$.
A policy $\pi = \{\mu_1, \mu_2, \ldots\}$ is referred to as
stationary if there is a function $\mu:{\cal Z}\mapsto {\cal U}$
such that $\mu_t(h_t) = \mu(z_{t-1})$ for all $t$ and $h_t$.  With
some abuse of terminology, we will sometimes refer to such a
function $\mu$ itself as a stationary policy.

We consider an objective of maximizing average reward, given a
bounded reward function $g:{\cal Z}\times{\cal U} \rightarrow
\Re$.  The average reward for a policy $\pi$ is defined by
$$\rho_\pi = \liminf_{N \to \infty} \frac{1}{N} \mathbb{E}_{\pi}\left\{\sum_{t=0}^{N-1}
g(Z_t, \mu_{t+1}(h_{t+1}))\right\},$$ where the subscript $\pi$
indicates that actions are generated by the policy $\pi =
(\mu_1,\mu_2,\ldots)$.  The optimal average reward is defined by
$$\rho^* = \sup_\pi \rho_\pi.$$

\subsection{The Bellman Equation}

An alternative characterization of the optimal average reward is
offered by the Bellman Equation.  This equation offers a mechanism
for verifying that a given level of average reward is optimal.  It
also leads to a characterization of optimal policies.  The
following result which we will later use encapsulates the Bellman
equation and its relation to the optimal average reward and
optimal policies.
\begin{theorem}
\label{th:Bellman} If $\rho \in \Re$ and a bounded function
$h:{\cal Z}\mapsto \Re$ satisfy
\begin{equation}
\label{eq:Bellman} \rho + h(z) = \sup_{u \in {\cal U}}
\left(g(z,u) + \int P_w(dw | z,u) h(F(z,u,w))\right) \: \: \forall
z\in{\cal Z}
\end{equation}
then $\rho=\rho^*$.  Further, if there is a function $\mu:{\cal
Z}\mapsto{\cal U}$ such that $\mu(z)$ attains the supremum for
each $z$ then $\rho_\pi = \rho^*$ for $\pi = (\mu_0,
\mu_1,\ldots)$ with $\mu_t(h_t) = \mu(z_{t-1})$ for each $t$.
\end{theorem}
\noindent This result follows immediately from Theorem 6.2 of
\cite{Arapos93_average_cose_survey}. It is convenient to define a
dynamic programming operator $T$ by
$$(Th)(z) = \sup_{u \in {\cal U}} \left(g(z,u) + \int P_w(dw | z,u) h(F(z,u,w))\right),$$
for all functions $h$.  Then, Bellman's equation can be written as
$\rho {\bf 1} + h = Th$. It is also useful to define for each
stationary policy $\mu$ an operator
$$(T_\mu h)(z) = g(z,\mu(z)) + \int P_w(dw | z,\mu(z)) h(F(z,\mu(z),w)).$$

The operators $T$ and $T_\mu$ obey some well-known properties.
First, they are monotonic: for bounded functions $h$ and
$\overline{h}$ such that $h \leq \overline{h}$, $Th \leq
T\overline{h}$ and $T_\mu h \leq T_\mu \overline{h}$.  Second,
they are non-expansive with respect to the sup-norm: for bounded
functions $h$ and $\overline{h}$, $\|T h - T \overline{h}\|_\infty
\leq \|h - \overline{h}\|_\infty$ and $\|T_\mu h - T_\mu
\overline{h}\|_\infty \leq \|h - \overline{h}\|_\infty$. Third, as
a consequence of nonexpansiveness, $T$ is continuous with respect
to the sup-norm.\footnote{The proof of the properties of $T$ are
entirely analogous to the proofs of Propositions 1.2.1 and 1.2.4
in \cite[Vol. II]{Bertsekas05}}

\subsection{Feedback Capacity as a Dynamic Program}
\label{se:unifilar-DP}

We will now formulate a dynamic program such that the optimal
average reward equals the feedback capacity of a unifilar channel
as presented in Theorem \ref{t_feedback_capacity_unifilar_FSC1}.
This entails defining the septuple $({\cal Z}, {\cal U}, {\cal W},
F, P_z, P_w, g)$ based on properties of the unifilar channel and
then verifying that the optimal average reward is equal to the
capacity of the channel.

Let $\beta_t$ denote the $|{\cal S}|$-dimensional vector of
channel state probabilities given information available to the
decoder at time $t$.  In particular, each component corresponds to
a channel state $s_t$ and is given by $\beta_t(s_t) \triangleq
p(s_t | y^t)$. We take states of the dynamic program to be $z_t =
\beta_t$.  Hence, the state space ${\cal Z}$ is the $|{\cal
S}|$-dimensional unit simplex. Each action $u_t$ is taken to be
the matrix of conditional probabilities of the input $x_t$ given
the previous state $s_{t-1}$ of the channel.  Hence, the action
space ${\cal U}$ is the set of stochastic matrices of dimension
$|{\cal S}|\times|{\cal X}|$. The disturbance $w_t$ is taken to be
the channel output $y_t$.  The disturbance space ${\cal W}$ is the
output alphabet ${\it Y}$.

The initial state distribution $P_z$ is concentrated at the prior
distribution of the initial channel state $s_0$.  Note that the
channel state $s_t$ is conditionally independent of the past given
the previous channel state $s_{t-1}$, the input probabilities
$u_t$, and the current output $y_t$.  Hence, $\beta_t(s_t) = p(s_t
| y^t) = p(s_t | \beta_{t-1}, u_t, y_t)$. More concretely, given a
policy $\pi = (\mu_1, \mu_2, \ldots)$,
\begin{eqnarray}\label{e_beta_iteration}
\beta_{t}(s_{t})
&=& p(s_t|y^t) \nonumber \\
&=& \sum_{x_{t},s_{t-1}} p(s_{t},s_{t-1},x_{t}|y^{t}) \nonumber \\
&=& \sum_{x_{t},s_{t-1}} \frac{p(s_{t},s_{t-1},x_{t},y_{t}|y^{t-1})}{p(y_{t}|y^{t-1})} \nonumber \\
&=& \sum_{x_{t},s_{t-1}}  \frac{p(s_{t-1}|y^{t-1})p(x_{t}|s_{t-1},y^{t-1})p(y_{t}|s_{t-1},x_{t})p(s_{t}|s_{t-1},x_{t},y_{t})}{p(y_{t}|y^{t-1})} \nonumber \\
& = & \frac{\sum_{x_{t},s_{t-1}}  \beta_{t-1}(s_{t-1})p(x_{t}|s_{t-1},y^{t-1})p(y_{t}|s_{t-1},x_{t})p(s_{t}|s_{t-1},x_{t},y_{t})}{\sum_{x_{t},s_{t},s_{t-1}} \beta_{t-1}(s_{t-1})p(x_{t}|s_{t-1},y^{t-1})p(y_{t}|s_{t-1},x_{t})p(s_{t}|s_{t-1},x_{t},y_{t})} \nonumber \\
& = & \frac{\sum_{x_{t},s_{t-1}} \beta_{t-1}(s_{t-1}) u_t(s_{t-1},
x_t) p(y_{t}|s_{t-1},x_{t})\mathbf{1}(s_t =
f(s_{t-1},x_{t},y_{t}))} {\sum_{x_{t},s_{t},s_{t-1}}
\beta_{t-1}(s_{t-1}) u_t(s_{t-1},x_t)
p(y_{t}|s_{t-1},x_{t})\mathbf{1}(s_t = f(s_{t-1},x_{t},y_{t}))},
\end{eqnarray}
where $\mathbf{1}(\cdot)$ is the indicator function. Note that
$p(y_t | s_{t-1},x_t)$ is given by the channel model. Hence,
$\beta_t$ is determined by $\beta_{t-1}$, $u_t$, and $y_t$, and
therefore, there is a function $F$ such that $z_t = F(z_{t-1},
u_t, w_t)$.

The distribution of the disturbance $w_t$ is
$p(w_t|z^{t-1},w^{t-1},u^t) = p(w_t | z_{t-1},u_t)$. Conditional
independence from $z^{t-2}$ and $w^{t-1}$ given $z_{t-1}$ is due
to the fact that the channel output is determined by the previous
channel state and current input.  More concretely,
\begin{eqnarray}
p(w_t|z^{t-1},w^{t-1},u^t)&= &p(y_t | \beta^{t-1}, y^{t-1}, u^t)\nonumber \\
& = & \sum_{x_t,s_{t-1}}
p(y_{t},x_t,s_{t-1}|\beta^{t-1},y^{t-1},u^t) \nonumber \\
& = & \sum_{x_t,s_{t-1}}
p(s_{t-1}|\beta_{t-1},u_t)p(x_t|s_{t-1},\beta_{t-1},u_t)p(y_{t}|x_t,s_{t-1},\beta_{t-1},u_t)\nonumber \\
& = & \sum_{x_t,s_{t-1}}
p(s_{t-1},x_t,y_t|\beta_{t-1},u_t) \nonumber \\
& = & p(y_t|\beta_{t-1},u_t)\nonumber\\
 & = & p(w_t|z_{t-1},u_t).
\end{eqnarray}
Hence, there is a disturbance distribution $P_w(\cdot |
z_{t-1},u_t)$ that depends only on $z_{t-1}$ and $u_t$.

We consider a reward of $I(Y_t;X_t,S_{t-1} | y^{t-1})$. Note that
the reward depends only on the probabilities $p(x_t, y_t, s_{t-1}
| y^{t-1})$ for all $x_t$, $y_t$ and $s_{t-1}$.  Further,% since
%the initial state $\beta_0$ is deterministic, given a policy $\pi
%= (\mu_1, \mu_2, \ldots)$,
\begin{eqnarray}\label{e_pxtytst-1|yt-1}
p(x_t, y_t, s_{t-1} | y^{t-1})
%&=& p(x_t, y_t, s_{t-1} | h^{t-1}) \\
&=& p(s_{t-1} | y^{t-1}) p(x_t | s_{t-1}, y^{t-1}) p(y_t | x_t,
s_{t-1}) \nonumber \\
%&=& \beta_{t-1}(s_{t-1}) (\mu_t(h^{t-1}))(s_{t-1}, x_t) p(y_t|x_t,s_{t-1}) \\
&=& \beta_{t-1}(s_{t-1}) u_t(s_{t-1}, x_t) p(y_t|x_t,s_{t-1}).
\end{eqnarray}
Recall that $p(y_t|x_t,s_{t-1})$ is given by the channel model.
Hence, the reward depends only on $\beta_{t-1}$ and $u_t$.

 Given an initial state $z_0$ and a policy $\pi =
(\mu_1, \mu_2, \ldots)$, $u_t$ and $\beta_t$ are determined by
$y^{t-1}$. Further, $(X_t,S_{t-1},Y_t)$ is conditionally
independent of $y^{t-1}$ given $\beta_{t-1}$ as shown in
(\ref{e_pxtytst-1|yt-1}). Hence,
\begin{equation}\label{e_reward}
g(z_{t-1},u_t)=I(Y_t;X_t,S_{t-1} | y^{t-1}) = I(X_t,S_{t-1};Y_t |
\beta_{t-1}, u_t).
\end{equation}
%and can be written as $g(z_{t-1},u_t) =  I(Y_t;X_t,S_{t-1} |
%y^{t-1})$.
 It follows that the optimal average reward is
$$\rho^* = \sup_\pi
\liminf_{N \rightarrow \infty} \frac{1}{N}
\mathbb{E}_\pi\left[\sum_{t=1}^{N}I(X_t,S_{t-1};Y_t|Y^{t-1})\right]=C_{FB}.$$

The dynamic programming formulation that is presented here is an
extension of the formulation presented in \cite{Yang05} by Yang,
Kav\u{c}i\'{c} and Tatikonda. In \cite{Yang05} the formulation is
for channels with the property that the state is deterministically
determined by the previous inputs and here we allow the state to
be determined by the previous outputs and inputs.

\section{Solution for the Trapdoor
Channel}\label{s_solution_for_the_chemical}

The trapdoor channel presented in Section
\ref{s_channel_model_and_pre} is a simple example of a unifilar
FSC. In this section, we present an explicit solution to the
associated dynamic program, which yields the feedback capacity of
the trapdoor channel as well as an optimal encoder-decoder pair.
The analysis begins with a computational study using numerical
dynamic programming techniques.  The results give rise to
conjectures about the average reward, the differential value
function, and an optimal policy.  These conjectures are proved to
be true through verifying that they satisfy Bellman's equation.

\subsection{The Dynamic Program}
\label{se:chemical-DP}

In Section \ref{se:unifilar-DP}, we formulated a class of dynamic
programs associated with unifilar channels.  From here on we will
focus on the particular instance from this class that represents
the trapdoor channel.

Using the same notation as in Section \ref{se:unifilar-DP}, the
state $z_{t-1}$ would be the vector of channel state probabilities
$[p(s_{t-1}=0|y^{t-1}),p(s_{t-1}=1|y^{t-1})]$.  However, to
simplify notation, we will consider the state $z_t$ to be the
first component; that is, $z_{t-1} \triangleq
p(s_{t-1}=0|y^{t-1})$. This comes with no loss of generality --
the second component can be derived from the first since the pair
sum to one. The action is a $2 \times 2$ stochastic matrix
\begin{equation} \label{e_pxtst}
u_t = \left[
\begin{array}{cc}
p(x_t=0|s_t=0) & p(x_t=1|s_t=0)\\
p(x_t=0|s_t=1)  & p(x_t=1|s_t=1) \\
\end{array}
\right].
\end{equation}
The disturbance $w_t$ is the channel output $y_t$.

The state evolves according to $z_t = F(z_{t-1}, u_t, w_t)$, where
using relations from eq. (\ref{e_chemical_st},
\ref{e_beta_iteration}) and Table \ref{ta_prob_chem}, we obtain
the function $F$ explicity as
$$z_t = \left\{\begin{array}{ll}
\frac{z_{t-1} u_t(1,1)}{z_{t-1} u_t(1,1) + 0.5 z_{t-1} u_t(1,2)
+ 0.5 (1- z_{t-1}) u_t(2,1)} \quad &{\rm if}\ w_t = 0 \\
& \\
\frac{0.5(1- z_{t-1}) u_t(2,1) + 0.5 z_{t-1} u_t(1,2)} {0.5
(1-z_{t-1}) u_t(2,1) + 0.5 z_{t-1} u_t(1,2) + (1-z_{t-1})
u_t(2,2)} \quad &{\rm if}\ w_t = 1.
\end{array}\right.$$
These expressions can be simplified by defining
\begin{equation}\label{e_gamma}
 \gamma_t
\triangleq (1-z_{t-1}) u_t(2,2),
\end{equation}
\begin{equation}\label{e_delta}
\delta_t \triangleq z_{t-1} u_t(1,1).
\end{equation}
 So that
$$z_{t} = \left\{\begin{array}{ll}
\frac{2 \delta_t}{1 + \delta_t - \gamma_t}  \quad & {\rm if}\ w_t = 0 \\
& \\
1 - \frac{2 \gamma_t}{1 - \delta_t + \gamma_t} & {\rm if}\ w_t =
1.
\end{array}\right.$$
Note that, given $z_{t-1}$, the action $u_t$ defines the pair
$(\gamma_t,\delta_t)$ and vice-versa. From here on we will
represent the action in terms of $\gamma_t$ and $\delta_t$.
Because $u_t$ is required to be a stochastic matrix, $\delta_t$
and $\gamma_t$ are constrained by $0\leq\delta_t\leq z_t$ and
$0\leq\gamma_t\leq 1- z_t$.

Recall from eq. (\ref{e_reward}) that the reward function is given
by $g(z_{t-1},u_t) = I(X_t,S_{t-1};Y_t | \beta_{t-1}, u_t)$. %(see .
%Given an initial state $z_0$ and a policy $\pi =
%(\mu_1, \mu_2, \ldots)$, $u_t$ and $\beta_t$ are determined by
%$y^{t-1}$. Further, $(X_t,S_{t-1},Y_t)$ is conditionally
%independent of $y^{t-1}$ given $\beta_{t-1}$.  Hence,
%$g(z_{t-1},u_t) = I(X_t,S_{t-1};Y_t | \beta_{t-1}, u_t)$.  Given
%$z_{t-1}$ and $u_t$,
This reward can be computed from the conditional probabilities
$p(x_t,s_{t-1},y_t | \beta_{t-1}, u_t)$. Using the expressions for
these conditional probabilities provided in Table
\ref{ta_dist_xtstyt}, we obtain
\begin{eqnarray*}
g(z_{t-1},u_t)
&=& I(X_t,S_{t-1};Y_t | \beta_{t-1}, u_t) \\
&=& H(Y_t | u_t, \beta_{t-1}) - H(Y_t | X_t,S_{t-1}, \beta_{t-1}, u_t) \\
&=& H\left(z_{t-1} u_t(1,1) + \frac{z_{t-1} u_t(1,1)}{2} +
\frac{(1-z_{t-1}) u_t(2,1)}{2}\right)
- z_{t-1} u_t(1,2) - (1-z_{t-1}) u_t(1,1) \\
&=& H\left(\frac{1}{2} + \frac{\delta_t-\gamma_t}{2}\right) +
\delta_t + \gamma_t - 1,
\end{eqnarray*}
where, with some abuse of notation, we use $H$ to denote the
binary entropy function: $H(q) = - q \ln q - (1-q) \ln(1-q)$.
%{\bf [Haim: I did not check this algebra.  I think its worth
%adding steps between the second and third that are easy to
%verify.]}

\begin{table}[h]
\caption{The conditional distribution $p(x_t,s_{t-1},y_t |
\beta_{t-1}, u_t)$.} \centering
\begin{tabular}[h]{|c|c|c|c|}
\hline
$x_t$ & $s_{t-1}$ & $y_t=0$ & $y_t=1$\\
\hline \hline
0 & 0 & $\beta_t u_t(1,1) $ & 0 \\
\hline
0 & 1 & $0.5(1-\beta_t) u_t(2,1) $ & $ 0.5 (1-\beta_t) u_t(2,1)$ \\
\hline
1 & 0 & $0.5 \beta_t u_t(1,2)$ & $0.5 \beta_t u_t(1,2)$\\
\hline
1 & 1 & 0 & $(1-\beta_t) u_t(1,2)$ \\
\hline
\end{tabular}
\label{ta_dist_xtstyt}
\end{table}

We now have a dynamic program -- the objective is to maximize over
all policies $\pi$ the average reward $\rho_\pi$.  The capacity of
the trapdoor channel is the maximum of the average reward
$\rho^*$.  In the context of the trapdoor channel, the dynamic
programming operator takes the form
\begin{equation}\label{e_belman}
(T h)(z) = \sup_{0 \leq \delta \leq z, 0\leq \gamma \leq 1-z}
\left(H\left(\frac{1}{2} +
\frac{\delta-\gamma}{2}\right)+\delta+\gamma-1
+\frac{1+\delta-\gamma}{2} h\left(\frac{2\delta}{
1+\delta-\gamma}\right)+\frac{1-\delta+\gamma}{2}h\left(1-\frac{2\gamma}{1-\delta+\gamma}
\right) \right).
\end{equation}
By Theorem \ref{th:Bellman}, if we identify a scalar $\rho$ and
bounded function $h$ that satisfy Bellman's equation, $\rho{\bf 1}
+ Th = h$, then $\rho$ is the optimal average reward.  Further, if
for each $z$, $T_\mu h = T h$ then the stationary policy $\mu$ is
an optimal policy.

\subsection{Computational Study}

We carried out computations to develop an understanding of
solutions to Bellman's equation. For this purpose, we used the
{\it value iteration} algorithm, which in our context generates a
sequence of iterates according to
\begin{equation}
\label{eq:value-iteration} J_{k+1} = T J_k,
\end{equation}
initialized with $J_0 = 0$.  For each $k$ and $z$, $J_k(z)$ is the
maximal expected reward over $k$ time periods given that the
system starts in state $z$.  Since rewards are positive, for each
$z$, $J_k(z)$ grows with $k$.  For each $k$, we define a
differential reward function $h_k(z) \triangleq J_k(z) - J_k(0)$.
These functions capture differences among values $J_k(z)$ for
different states $x$.  Under certain conditions such as those
presented in \cite{ZhuGuo2005}, the sequence $h_k$ converges
uniformly to a function that solves Bellman's equation. We will
neither discuss such conditions nor verify that they hold. Rather,
we will use the algorithm heuristically in order to develop
intuition and conjectures.

Value iteration as described above cannot be implemented on a
computer because it requires storing and updating a function with
infinite domain and optimizing over an infinite number of actions.
To address this, we discretize the state and action spaces,
approximating the state space using a uniform grid with $2000$
points in the unit interval and restricting actions $\delta$ and
$\gamma$ to values in a uniform grid with $4000$ points in the
unit interval.

\begin{figure}[h]{
\psfrag{relative frequence}[][][0.9]{relative frequency}
\psfrag{Value function}[][][0.9]{Value function on the $20^{\text
th}$ iteration, $J_{20}$} \psfrag{J20}[][][1]{$J_{20}$}
\psfrag{action-parameter delta}[][][0.9]{Action-parameter,
$\delta$} \psfrag{action-parameter
gamma}[][][0.9]{Action-parameter, $\gamma$} \psfrag{histogram of
beta}[][][0.9]{Histogram of $z$}
 \psfrag{delta}[][][1]{$\delta$}
 \psfrag{gamma}[][][1]{$\gamma$}
\psfrag{beta}[][][1]{$z$}

\centerline{\includegraphics[width=12cm]{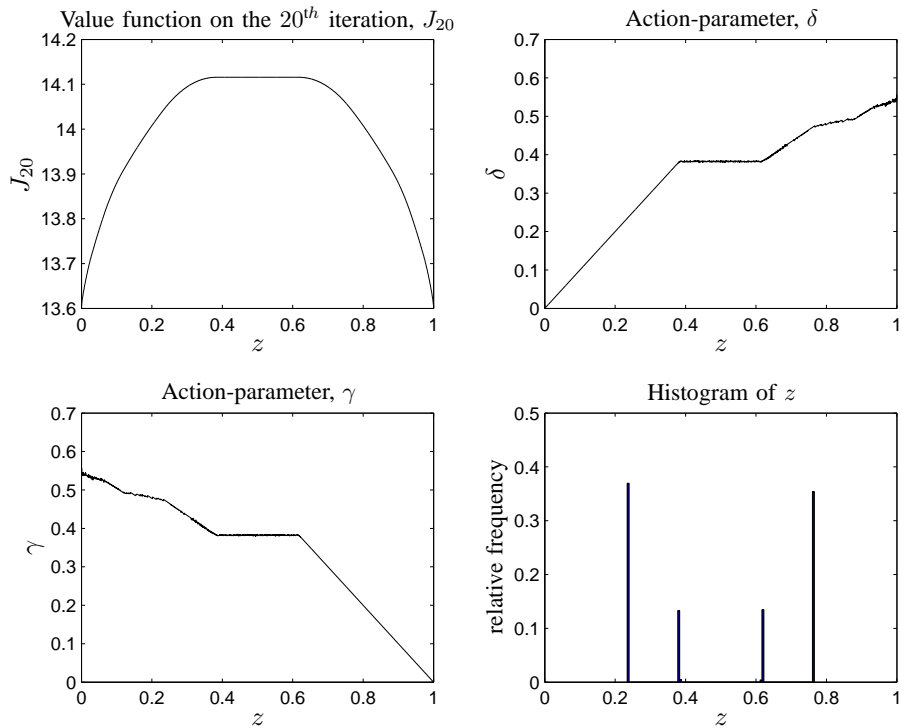}}
%\centerline{\includegraphics[width=9cm]{C:/haim/matlab/chemical2/f_iteration_20_all.eps}}
\caption{Results from 20 value iterations. On the top-left side
the value function $J_{20}$ is plotted. On the top-right and
bottom-left the optimal action-parameters $\delta$ and $\gamma$
with respect to $20^{\text th}$ iteration are plotted. On the
bottom-right the relative state frequencies of the associated
Markov process of $z$ with the policy that is optimal with respect
to $J_{20}$ is plotted.\label{f_res_value}}
}\end{figure}

We executed twenty value iterations.  Figure \ref{f_res_value}
plots the function $J_{20}$ and actions that maximize the
right-hand-side of eq. (\ref{eq:value-iteration}) with $k=20$. We
also simulated the system, selecting actions $\delta_t$ and
$\gamma_t$ in each time period to maximize this expression.  This
led to an average reward of approximately $0.694$.  We plot in the
right-bottom side of Figure \ref{f_res_value} the relative state
frequencies of the associated Markov process.  Note that the
distribution concentrates around four points which are
approximately $0.236$, $0.382$,  $0.613$, and $0.764$.

 %%when I save on visio I used AI 3 - and I had to restart the computer and not all visio features work with AI3

\subsection{Conjectures} \label{s_conjecture}

The results obtained from value iteration were, amazingly, close
to the answers of two questions given in an information theory
class at Stanford taught by Professor Thomas Cover. Here is a
simplified version of the questions given to the class.
\begin{itemize}
\item [(1)]{\it Entropy rate.} Find the maximum entropy rate of the two-state Markov
chain (Fig. \ref{f_markov_hw}) with transition matrix
\begin{equation}
P=\left[ \begin{array}{cc} 1-p & p \\ 1 & 0\end{array} \right],
\end{equation}
where $0\leq p \leq 1$ is the free parameter we maximize over.

\begin{figure}[h]{
\psfrag{p}[][][1]{$p$} \psfrag{1-p}[][][1]{$1-p$}
\psfrag{1}[][][1]{$1$} \psfrag{A}[][][1]{$0$}
\psfrag{B}[][][1]{$1$}

\centerline{\includegraphics[width=4cm]{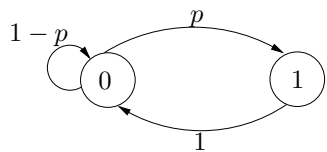}}
%\centerline{\includegraphics[width=4cm]{C:/cygwin/usr/X11R6/bin/home/markov_hw.eps}}
\caption{The Markov chain of question 1.} \label{f_markov_hw} }
\end{figure}

\item [(2)]
{\it Number of sequences. } To first order in the exponent, what
is the number of binary sequences of length $n$ with no two 1's in
a row?
\end{itemize}

 The entropy rate of the Markov chain of question (1) is given
by $\frac{H(p)}{1+p}$, and when maximizing over $0\leq p \leq 1$,
we get that $p=\frac{3-\sqrt 5}{2}$ and the entropy rate is
0.6942.  It can be shown that the number of sequences of length
$n-1$ that do not have two 1's in a row is the $n^{th}$ number in
the Fibonacci sequence. This can be proved by induction in the
following way. Let us denote ($N^0_n,N^1_n$) the number of
sequences of length $n$ with the condition of not having two 1's
in a row that are ending with `0' and with `1' respectively. For
the sequences that end with `0' we can either add a next bit `1'
or `0' and for the sequences that end with `1' we can add only
`0'. Hence $N^0_{n+1}=N^0_{n}+N^1_{n}$ and $N^1_{n+1}=N^0_{n}$. By
repeating this logic, we get that $N^0_{n}$ behaves as a Fibonacci
sequence.
 To first order in the exponent, the
Fibonacci number behaves as $\lim_{n\to \infty} \frac{1}{n}\log
f_n =\log \frac{1+\sqrt5}{2}=0.6942$, where the number,
$\frac{1+\sqrt5}{2}$, is called the golden ratio. The golden ratio
is also known to be a positive number that solves the equation
$\frac{1}{\phi}=1-\phi$, and it appears in many math, science and
art problems \cite{Mario_golden_ratio}. As these problems
illustrate, the number of typical sequences created by the Markov
process given in question (1) is, to first order in the exponent,
equal to the number of binary sequences that do not have two 1's
in a row.

Let us consider a policy for the dynamic program associated with a
binary random process that is created by the Markov chain from
question 1 (see Fig \ref{f_markov_hw}). Let the state of the
Markov process indicate if the input to the channel will be the
same or different from the state of the channel. In other words,
if at time $t$ the binary Markov sequence is `0' then the input to
the channel is equal to the state of the channel, i.e.
$x_t=s_{t-1}$. Otherwise, the input to the channel is a complement
to the state of the channel, i.e. $x_t=s_{t-1}\oplus1$. This
scheme uniquely defines the distribution $p(x_t|s_{t-1},y^{t-1})$:

\begin{equation}
p(X_t=s_{t-1}|s_{t-1},y_{t-1})=\left\{ \begin{array}{ll} 1-p&
\text{ if } s_{t-1}=y_{t-1},\\ 1 & \text{ if } s_{t-1}\neq
y_{t-1}.
\end{array} \right.
\end{equation}
This distribution is derived from the fact that for the trapdoor
channel the state evolves according to equation
(\ref{e_chemical_st}) which can be written as
\begin{equation}
s_{t-1}\oplus y_{t-1}=x_{t-1}\oplus s_{t-2}.
\end{equation}
Hence, if $s_{t-1}\neq y_{t-1}$ then necessarily also $x_{t-1}\neq
s_{t-2}$. This means that the tuple $(s_{t-1},y_{t-1})$ defines
the state of the Markov chain at time $t-1$ and the tuple
$(x_{t},s_{t-1})$ defines the state of the Markov chain at time
$t$. Having the distribution $p(x_t|s_{t-1},y^{t-1})$, for the
following four values of $z$, $\{b_1\triangleq \sqrt 5 -2,
b_2\triangleq \frac{3-\sqrt 5 }{2}, b_3\triangleq \frac{\sqrt 5
-1}{2}, b_4\triangleq 3-\sqrt 5 \}$, the corresponding actions
$\tilde{\gamma}(z)$ and $\tilde{\delta}(z)$ which are defined in
eq. (\ref{e_gamma},\ref{e_delta}) are:
\begin{center}
\begin{tabular}{|c|c|c|}
\hline $z$ & $\tilde{\gamma}(z)$ & $\tilde{\delta}(z)$ \\
\hline $b_1$ or $b_2$&$\frac{\sqrt 5-1 }{2} (1-z)$ & $z$ \\
%\hline $\frac{3-\sqrt 5 }{2}$&$\frac{3-\sqrt 5 }{2}(1-\beta)$&$\beta$\\
\hline $b_3$ or $b_4$ &$1-z$&$\frac{\sqrt 5-1 }{2}z$\\
%\hline $\frac{(3-\sqrt 5)}{2}$&$0$&$\frac{(\sqrt 5 -1)}{2}\beta$\\
\hline
\end{tabular}
\end{center}

\begin{figure}[h]{
\psfrag{Y1}[][][1]{$1$}\ \psfrag{y2}[][][1]{$1$}
\psfrag{y3}[][][1]{$1$}\ \psfrag{y4}[][][1]{$1$}
\psfrag{y5}[][][1]{$0$}\ \psfrag{y6}[][][1]{$0$}
\psfrag{y7}[][][1]{$0$}\ \psfrag{y8}[][][1]{$0$}
\psfrag{a}[][][1]{$b_1$}\ \psfrag{b}[][][1]{$b_2$}
\psfrag{c}[][][1]{$b_3$}\ \psfrag{d}[][][1]{$b_4$}
\psfrag{Bt}[][][1]{$\beta_{t-1}$}\ \psfrag{Ct}[][][1]{$\beta_{t}$}
\psfrag{Yt}[][][1]{$y_{t}$}

\centerline{\includegraphics[width=6cm]{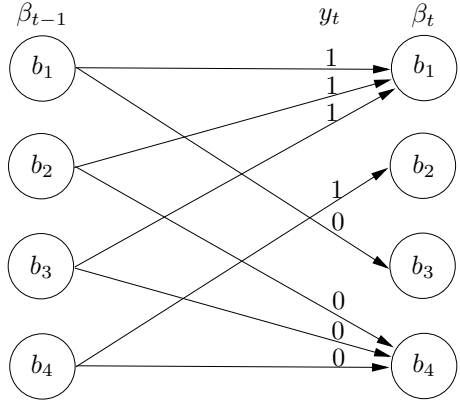}}
%\centerline{\includegraphics[width=6cm]{C:/cygwin/usr/X11R6/bin/home/beta_transition.eps}}
\caption{The transition between $\beta_{t-1}$ and $\beta_t$, under
the policy $\tilde \delta, \tilde
\gamma$.\label{f_beta_trensition}} }
\end{figure}

It can be verified, by using eq. (\ref{e_beta_iteration}), that
the only values of $z$ ever reached are,
\begin{equation}\label{e_4_states}
z \in \left\{b_1\triangleq \sqrt 5 -2, b_2\triangleq \frac{3-\sqrt
5 }{2}, b_3\triangleq \frac{\sqrt 5 -1}{2}, b_4\triangleq 3-\sqrt
5 \right\} ,
\end{equation}
%\begin{equation}\label{e_4_states}
%\beta \in \left\{\sqrt 5 -2, \frac{3-\sqrt 5 }{2}, \frac{\sqrt 5
%-1}{2}, 3-\sqrt 5 \right\} .
%\end{equation}
and the transitions are a function of $y_t$ shown graphically in
Figure \ref{f_beta_trensition}.
%
%Under the policy given here the transition between the four
%different states is a function of $y_t$ and it is described
%graphically in figure \ref{f_beta_trensition}.
Our goal is to prove that an extension of this policy is indeed
optimal.  Based on the result of Question 1, we conjugate that the
entropy rate of the average reward is
\begin{equation}\label{e_average_reward_hu}
\tilde{\rho}
=\frac{H(\frac{3-\sqrt5}{2})}{1+\frac{3-\sqrt5}{2}}=\log
\frac{\sqrt5+1}{2} \approx 0.6942.
\end{equation}
It is interesting to notice that all the numbers appearing above
can be written in terms of the golden ratio, $\phi =
\frac{\sqrt{5} - 1}{2}$.  In particular, $\tilde{\rho} = \log
\phi$, $b_1= 2 \phi-3$, $b_2=2-\phi$, $b_3=\phi-1$ and
$b_4=4-2\phi$.

By inspection of Figure \ref{f_res_value}, we let $\tilde{\gamma}$
and $\tilde{\delta}$ be linear over the intervals $[b_1, b_2]$,
$[b_2,b_3]$, and $[b_3,b_4]$ and we get the form presented in
Table \ref{t_opt_policy}.
\begin{table}
\begin{center}
\begin{tabular}{|c|c|c|}
\hline $z$ & $\tilde{\gamma}(z)$ & $\tilde{\delta}(z)$ \\
\hline $b_1 \leq z \leq b_2$&$\frac{\sqrt 5-1 }{2}(1-z)$ & $ z$ \\
\hline $b_2 \leq z \leq b_3$&$ \frac{3-\sqrt 5 }{2}$ & $\frac{3-\sqrt 5 }{2}$ \\
\hline $b_3 \leq z \leq  b_4$ &$1-z$&$\frac{\sqrt 5 -1}{2} z$\\
%\hline $\frac{(3-\sqrt 5)}{2}$&$0$&$\frac{(\sqrt 5 -1)}{2}\beta$\\
\hline
\end{tabular}
\end{center}
\caption{Conjectured policy which in the next section will be
proven to be true.\label{t_opt_policy}}
\end{table}
%We don't explicitly assign actions for states outside $[b_1,b_4]$.
%This interval is absorbing, any policy for which the probability
%of being absorbed converges to one at an exponential rate will do.
%Almost any policy satisfies this condition.

We now propose differential values $\tilde{h}(z)$ for $z \in [b_1,
b_4]$. If we assume that $\tilde{\delta}$ and $\tilde{\gamma}$
maximize the right-hand-side of the Bellman equation (eq.
\ref{eq:Bellman}) for $z \in [b_1,b_4]$ with $h = \tilde{h}$ and
$\rho = \tilde{\rho}$, we obtain
\begin{equation}\label{e_J2}
\tilde{h}(z)=H\left(\frac{1}{2}\right)-(\sqrt 5 -2) - \tilde{\rho}
+ \tilde{h}(3-\sqrt 5), \qquad b_2 \leq z \leq b_3,
\end{equation}
\begin{equation}\label{e_J3}
\tilde{h}(z)=H\left(\frac{\sqrt5 +1}{4}z\right)-\frac{3-\sqrt5}{2}
z - \rho+\frac{\sqrt5 +1}{4} z \tilde h(3-\sqrt
5)+\left(1-\frac{\sqrt5 +1}{4}z \right)\tilde
h\left(\frac{1-z}{1-\frac{\sqrt5 +1}{4} z} \right), \qquad b_3
\leq z \leq b_4.
\end{equation}
The equation for the range $b_1 \leq z \leq b_2$ is implied by the
symmetry relation:  $\tilde{h}(z)=\tilde{h}(1-z)$.

If a scalar $\rho$ and function $h$ solve Bellman's equation, so
do $\rho$ and $h + c {\bf 1}$ for any scalar $c$.  Therefore,
there is no loss of generality in setting $\tilde{h}(1/2) = 1$.
From eq. (\ref{e_J2}) we have that
\begin{equation}\label{e_Jb2b3}
\tilde{h}(z) = 1,\qquad b_2 \leq z \leq b_3.
\end{equation}
In addition, by symmetry considerations we can deduce that
$\tilde{h}(\sqrt 5 -2)=\tilde h(3-\sqrt 5)$ and from eq.
(\ref{e_J2}) we obtain
\begin{equation}
\tilde{h}(\sqrt 5 -2) = \tilde{h}(3-\sqrt 5) = \tilde{\rho} - 2 +
\sqrt 5 \approx 0.9303.
\end{equation}
Taking symmetry into consideration and applying eq. (\ref{e_J3})
twice we obtain,
\begin{equation}\label{e_Jb3b4}
\tilde{h}(z)=H(z)+ \tilde{\rho} z + c_1, \qquad b_3 \leq z \leq
b_4,
\end{equation}
where $c_1=\log(3-\sqrt 5)$. %=-0.3884$.
By symmetry we obtain
\begin{equation}\label{e_Jb1b2}
\tilde{h}(z)=H(z)- \tilde{\rho} z + c_2, \qquad b_1 \leq z \leq
b_2.
\end{equation}
where $c_2=\log(\sqrt 5  -1)$

\begin{figure}[h]{%\label{f_res_value}
\psfrag{g1}[][][0.8]{$1-z \longrightarrow$}
\psfrag{g2}[][][0.8]{$\frac{3-\sqrt 5}{2}$}
\psfrag{g3}[][][0.8]{$\frac{\sqrt 5 -1}{2} (1-z)$} \psfrag{g4}[][][0.8]{$\downarrow$}%\downarrow{$\searrow$}

\psfrag{d1}[][][0.8]{$\longleftarrow z$}
\psfrag{d2}[][][0.8]{$\frac{3-\sqrt 5 }{2}$}
\psfrag{d3}[][][0.8]{$\nwarrow\frac{\sqrt 5 -1}{2}z$}
%\psfrag{d4}[][][0.8]{$\searrow$}

\psfrag{j1}[][][0.8]{$H(z)-\tilde \rho z+c_2$}
\psfrag{j0}[][][0.8]{$\searrow$}

\psfrag{j2}[][][0.8]{$$} \psfrag{j3}[][][0.8]{$H(z)+\tilde \rho
z+c_1$} \psfrag{j4}[][][0.8]{$\swarrow$}
\psfrag{10}[][][0.8]{$b_1$}\psfrag{2}[][][0.8]{$\downarrow$}
\psfrag{3}[][][0.8]{$b_2$} \psfrag{4}[][][0.8]{$\downarrow$}
\psfrag{5}[][][0.8]{$b_3$} \psfrag{6}[][][0.8]{$\downarrow$}
\psfrag{7}[][][0.8]{$b_4$}\psfrag{8}[][][0.8]{$\downarrow$}

\psfrag{h}[][][1]{$\tilde h$} \psfrag{parameter-action
delta}[][][0.9]{Action-parameter, $\tilde \delta$}
\psfrag{parameter-action gamma}[][][0.9]{Action-parameter, $\tilde
\gamma$}
 \psfrag{delta}[][][1]{$\tilde \delta$}
 \psfrag{gamma}[][][1]{$\tilde \gamma$}
\psfrag{beta}[][][1]{$z$}

\centerline{\includegraphics[width=12cm]{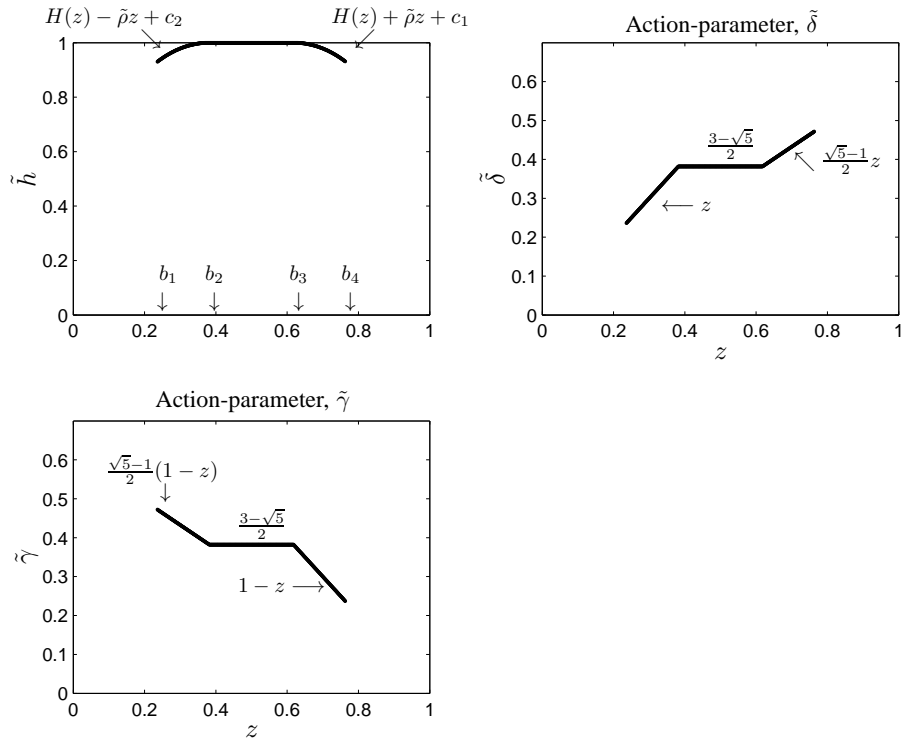}}
%\centerline{\includegraphics[width=9cm]{C:/haim/matlab/chemical2/guess_of_optimal_policy.eps}}
%\centerline{\includegraphics[width=16cm]{sd1.eps}}
\caption{A conjecture about the optimal solution based on the
$20^{\text th}$ value iteration of the DP which is shown in Fig.
\ref{f_res_value} and on the questions given by Professor Cover.
On the top-left the conjectured differential value $\tilde{h}(z)$
is plotted for $z\in [b_1,b_4]$. On the top-right side and
bottom-left the conjectured policy
$(\tilde{\delta}(z)$,$\tilde{\gamma}(z))$ is plotted for $z\in
[b_1,b_4]$} \label{f_n}
}\end{figure} %%when I save on visio I used AI 3 - and I had to restart the computer and not all visio features work with AI3

The conjectured policy $(\tilde \gamma, \tilde \delta)$, which is
given in Table \ref{t_opt_policy}, and the conjectured
differential value $\tilde h$, which is given in eq.
(\ref{e_Jb2b3})-(\ref{e_Jb1b2}), are plotted in Fig. \ref{f_n}.

\subsection{Verification}

In this section, we verify that the conjectures made in the
previous section are correct. Our verification process proceeds as
follows.  First, we establish that if a function $h:[0,1]\mapsto
\Re$ is concave, so is $Th$.  In other words, value iteration
retains concavity.  We then consider a version of value iteration
involving an iteration $h_{k+1} = T h_k - \tilde{\rho} {\bf 1}$.
Since subtracting a constant does not affect concavity, this
iteration also retains concavity.  We prove that if a function
$h_0$ is the pointwise maximum among concave functions that are
equal to $\tilde{h}$ in the interval $[b_1,b_4]$ then each iterate
$h_k$ is also concave and equal to $\tilde{h}$ in this interval.
Further, the sequence is pointwise nonincreasing.    These
properties of the sequence imply that it converges to a function
$h^*$ that again is concave and equal to $\tilde{h}$ in the
interval $[b_1,b_4]$.  This function $h^*$ together with
$\tilde{\rho}$ satisfies Bellman's Equation.  Given this, Theorem
\ref{th:Bellman} verifies our conjectures.

We begin with a lemma that will be useful in showing that value
iteration retains concavity.
\begin{lemma}\label{l_concavity}
Let $\zeta:[0,1]\times [0,1] \mapsto \Re$ be concave on
$[0,z]\times [0,1-z]$ for all $z \in [0,1]$ and
$$\psi(z)=\sup_{\delta \in [0,z], \gamma \in [0,1-z]} \zeta(\delta,\gamma).$$
Then $\psi:[0,1]\mapsto \Re$ is concave.
\end{lemma}
The proof of Lemma \ref{l_concavity} is given in the appendix.
%\ref{a_l_concavity}.
%We use the above lemma to establish that application of the
%operator $T$ retains concavity.

\begin{lemma} \label{le:concave} The operator $T$, defined in (\ref{e_belman})  retains
concavity and continuity. Namely,
\begin{itemize}
\item if $h$ is concave then $Th$ is concave,
\item if $h$ is continuous then $Th$ is continuous.
\end{itemize}
\end{lemma}

\emph{Proof (concavity):}
 It is well-known that the binary entropy function $H$ is concave, so the reward function
$$H\left(\frac{1}{2} + \frac{\delta-\gamma}{2}\right)+\delta+\gamma-1$$
is concave in $(\delta,\gamma)$.

Next, we show that if $h(z)$ is concave then
$\frac{1+\delta-\gamma}{2} h\left(\frac{2\delta}{
1+\delta-\gamma}\right)$ is concave in $(\delta,\gamma)$. Let
$\xi_1=\frac{1+\delta_1-\gamma_1}{2}$ and
$\xi_2=\frac{1+\delta_2-\gamma_2}{2}$.  We will show that, for any
$\alpha \in (0,1)$,
\begin{equation}
\alpha \xi_1 h\left(\frac{\delta_1}{\xi_1}\right)+ (1-\alpha)
\xi_2 h\left(\frac{\delta_2}{\xi_2}\right)\geq
(\alpha\xi_1+(1-\alpha)\xi_2)h\left(\frac{\alpha \delta_1 +
(1-\alpha)\delta_2}{\alpha \xi_1 +(1-\alpha)\xi_2}\right).
\end{equation}
Dividing both sides by $(\alpha\xi_1+(1-\alpha)\xi_2)$ we get
\begin{equation}
\frac{\alpha \xi_1}{\alpha\xi_1+(1-\alpha\xi_2)}
h\left(\frac{\delta_1}{\xi_1}\right)+ \frac{(1-\alpha)
\xi_2}{\alpha\xi_1+(1-\alpha\xi_2)}
h\left(\frac{\delta_2}{\xi_2}\right)\geq h\left(\frac{\alpha
\delta_1 + (1-\alpha)\delta_2}{\alpha \xi_1
+(1-\alpha)\xi_2}\right).
\end{equation}
Note that the last inequality is true because of the concavity of
$h$. It follows that
\begin{equation}\label{e_fconcave}
f(\delta,\gamma) \triangleq H\left(\frac{1}{2} +
\frac{\delta-\gamma}{2}\right)+\delta+\gamma-1+\frac{1+\delta-\gamma}{2}
h\left(\frac{2\delta}{
1+\delta-\gamma}\right)+\frac{1-\delta+\gamma}{2}
h\left(1-\frac{2\gamma}{1-\delta+\gamma} \right)
\end{equation}
 is concave in
$(\delta,\gamma$).  Since
$$(T h)(z) = \sup_{\delta \in [0,z], \gamma \in [0,1-z]} f(\delta,\gamma),$$
it is concave by Lemma \ref{l_concavity}. \hfill \QED

%The operator $T$ also retains continuity, as established in the
%following lemma.
%\begin{lemma}
%\label{le:continuous} If $h$ is continuous then $Th$ is
%continuous.
%\end{lemma}
\emph{Proof (continuity):} Note that the binary entropy function
$H$ is continuous. Further,
$h\left(\frac{2\delta}{1+\delta-\gamma}\right)$ and $ \quad
h\left(1-\frac{2\gamma}{1-\delta+\gamma} \right),$ are continuous
over the region $\{(\delta,\gamma)|\delta \geq 0, \gamma \geq 0,
\delta+\gamma \leq 1\}$. It follows that
%$$\zeta(\delta,\gamma) = H\left(\frac{1}{2} +
%\frac{\delta-\gamma}{2}\right)+\delta+\gamma-1+\frac{1+\delta-\gamma}{2}
%h\left(\frac{2\delta}{
%1+\delta-\gamma}\right)+\frac{1-\delta+\gamma}{2}
%h\left(1-\frac{2\gamma}{1-\delta+\gamma} \right)$$
$f(\delta,\gamma)$ is continuous over the region
$\{(\delta,\gamma)|\delta \geq 0, \gamma \geq 0, \delta+\gamma
\leq 1\}$.  Hence,
$$(Th)(z) = \sup_{\delta \in [0,z], \gamma \in  [0,1-z]} f(\delta,\gamma)$$
is continuous over $[0,1]$. \hfill \QED

%
%{\it Value iteration function $h_k(z)$:} We construct the value
%iteration in a way that will allow us proving uniformly
%convergence of the function $h_k(z)$. Let $h_0(z)$ be the
%pointwise maximum among concave functions satisfying $h_0(z) =
%\tilde{h}(z)$ for $z \in [b_1,b_4]$, where $\tilde{h}(z)$ is
%defined in eq.(\ref{e_Jb2b3})-(\ref{e_Jb1b2}). Let
%\begin{equation}
%h_{k+1}(z) = (T h_k)(z) - \tilde{\rho}.
%\end{equation}

Let us construct {\it value iteration function $h_k(z)$} as
follows. Let $h_0(z)$ be the pointwise maximum among concave
functions satisfying $h_0(z) = \tilde{h}(z)$ for $z \in
[b_1,b_4]$, where $\tilde{h}(z)$ is defined in
eq.(\ref{e_Jb2b3})-(\ref{e_Jb1b2}). Note that $h_0(z)$ is concave
and that for $z \notin [b_1,b_4]$, $h(z)$ is a linear
extrapolation from the boundary of $[b_1,b_4]$. Let
\begin{equation}\label{e_hk+1def}
h_{k+1}(z) = (T h_k)(z) - \tilde{\rho},
\end{equation}
and
\begin{equation}h^*(z)\triangleq  \limsup_{k\to \infty} h_k(z).
\end{equation}

The following lemma shows several properties of the sequence of
function $h_k(z)$ including the uniform convergence. The uniform
convergence is needed for verifying the conjecture, while the
other properties are intermediate steps in proving the uniform
convergence.

\begin{lemma}\label{l_hk}
The following properties hold:
\begin{itemize}
\item [{\ref{l_hk}.1}] for all $k\geq 0$, $h_k(z)$ is concave and continuous in $z$

\item [{\ref{l_hk}.2}] for all $k\geq 0$, $h_k(z)$ is symmetric
around $\frac{1}{2}$, i.e.
\begin{equation}
h_k(z)=h_k(1-z)
\end{equation}

\item [{\ref{l_hk}.3}]
 for all $k\geq 0$, $h_k(z)$ is fixed point for $z \in [b_1,b_4]$,
 i.e.,
\begin{equation}
h_k(z) = \tilde{h}(z), \qquad  z \in [b_1,b_4],
\end{equation}
and the stationary policy $\mu(z) = (\tilde{\delta}(z),
\tilde{\gamma}(z))$, where $(\tilde{\delta}(z),
\tilde{\gamma}(z))$ are defined in Table \ref{t_opt_policy},
satisfies $(T_\mu h_k)(z) = (T h_k)(z)$

\item [{\ref{l_hk}.4}] $h_k(z)$ is uniformly bounded in $k$ and $z$,
 i.e.,
\begin{equation}
\sup_k \sup_{z \in [0,1]} |h_k(z)| < \infty
\end{equation}

\item [{\ref{l_hk}.5}]
$h_k(z)$ is monotonically nonincreasing in $k$, i.e.
\begin{equation} \lim_{k\to \infty}
h_k(z)=h^*(z)
\end{equation}
\item [{\ref{l_hk}.6}]
$h_k(z)$ converges uniformly to $h^*(z)$
\end{itemize}

\end{lemma}

{\it Proof of }{\ref{l_hk}.1}: Since $h_0(z)$ is concave and
continuous and since the operator $T$ retain continuity and
concavity (see Lemma \ref{le:concave}),it follows that $h_k(z)$ is
concave and continuous for every $k$.\hfill \QED

{\it Proof of }{\ref{l_hk}.2}: We prove this property by
induction. First notice that $h_0(z)$ is symmetric and satisfies
$h_0(z)=h_0(1-z)$. Now let us show that if it holds for $h_k$ then
it holds for $h_{k+1}$.

Let $f_k(\delta, \gamma)$ denote the expression maximized to
obtain $(Th_k)(z)$, i.e.
\begin{equation}\label{e_fconcave}
f_k(\delta,\gamma) \triangleq H\left(\frac{1}{2} +
\frac{\delta-\gamma}{2}\right)+\delta+\gamma-1+\frac{1+\delta-\gamma}{2}
h_k\left(\frac{2\delta}{
1+\delta-\gamma}\right)+\frac{1-\delta+\gamma}{2}
h_k\left(1-\frac{2\gamma}{1-\delta+\gamma} \right).
\end{equation}
Notice that $f_k(\delta,\gamma)=f_k(\gamma,\delta)$. Also observe
that replacing the argument $z$ with $1-z$ in $Th_k$ yield the
same result as exchanging between $\gamma$ and $\delta$. From
those two observations follows that $Th_k(z)=Th_k(1-z)$ and from
the definition of $h_{k+1}$ given in (\ref{e_hk+1def}) follows
that $h_{k+1}(z)=h_{k+1}(1-z)$.\hfill \QED

 {\it Proof of }{\ref{l_hk}.3}: We prove this property by induction. Notice that
$h_0$ satisfies $h_0(z)=\tilde h(z)$ for $z\in[b_1, b_4]$. We
assume that $h_k$ satisfies $h_k(z)=\tilde h(z)$ and then we will
prove the property for $h_{k+1}$. We will show later in this proof
that for $z \in [b_1,b_4]$,
\begin{equation}\label{e_Thk_equal}
(T_\mu h_k)(z) = (T h_k)(z).
\end{equation}
Since $(T_\mu h_{k})(z) - \tilde{\rho} = \tilde h(z)$ for all $z
\in [b_1,b_4]$ (see eq.\ref{e_J2}-\ref{e_Jb1b2} ) it follows that
$h_{k+1}(z)=\tilde h(z)$ for all $z \in [b_1,b_4]$.

Now, let us show that (\ref{e_Thk_equal}) holds. Recall that in
the proof of Lemma \ref{le:concave}, eq. (\ref{e_fconcave}), we
showed that $f_k(\delta, \gamma)$ is concave in $(\delta,
\gamma)$. The derivative with respect to $\delta$ is,
\begin{eqnarray}
\frac{\partial f_k(\delta,\gamma)}{\partial \delta} &=&
\frac{1}{2}\log
\frac{1-\delta+\gamma}{1+\delta-\gamma}+1+\frac{1}{2}h_k\left(\frac{2\delta}{1+\delta-\gamma}\right)
-\frac{1}{2}h_k\left(\frac{2\gamma}{1-\delta+\gamma}\right)
\nonumber
\\&&
+\frac{1-\gamma}{1+\delta-\gamma}h'_k\left(\frac{2\delta}{1+\delta-\gamma}\right)
+\frac{\gamma}{1-\delta+\gamma}h'_k\left(\frac{2\gamma}{1-\delta+\gamma}\right).
\end{eqnarray}
The derivative with respect to $\gamma$ is entirely analogous and
can be obtained by mutually exchanging $\gamma$ and $\delta$.

For $z \in [b_2, b_3]$, the action $\tilde \gamma(z)=\tilde
\delta(z)=\frac{3-\sqrt 5 }{2}$ is feasible and $\frac{2
\tilde{\gamma}(z)}{1-\tilde{\delta}(z)+\tilde{\gamma}(z)} =
\frac{2\tilde{\delta}(z)}{1+\tilde{\delta}(z)-\tilde{\gamma}(z)} =
b_4$. Moreover, it is straightforward to check that the
derivatives of $f_k$ are zero at $(\tilde\gamma(z),\tilde
\delta(z))$, and since $f_k$ is concave, $(\tilde \gamma(z),
\tilde \delta(z))$ attains the maximum.  Hence, $(T_\mu h)(z) = (T
h)(z)$ for $z \in [b_2, b_3]$.

For $z \in [b_3,b_4]$, $\tilde \gamma(z) = 1-z$ and $\tilde
\delta(z) = \frac{\sqrt5-1}{2}z$. Note that
$\frac{2\tilde{\gamma}(z)}{1-\tilde{\delta}(z)+\tilde{\gamma}(z)}$
and
$\frac{2\tilde{\delta}(z)}{1+\tilde{\delta}(z)-\tilde{\gamma}(z)}$
are in $[b_1,b_2]\cup[b_3,b_4]$. Using expressions for
$\tilde{h}(z)$ given in equations (\ref{e_Jb3b4}) and
(\ref{e_Jb1b2}), we can write derivatives of $f$ at
$(\tilde{\delta}(z),\tilde{\gamma}(z))$ as
\begin{equation}
\frac{\partial f(\tilde{\delta}(z),\tilde{\gamma}(z))}{\partial
\delta} = \log \frac{1-\tilde{\delta}(z)-\tilde{\gamma}(z)}{2
\tilde{\delta}(z)} + 1 + \tilde\rho=0,
\end{equation}
\begin{equation}
\frac{\partial f(\tilde{\delta}(z), \tilde{\gamma}(z))}{\partial
\gamma} = \log \frac{1- \tilde{\delta}(z) - \tilde{\gamma}(z)}{2
\tilde{\gamma}(z)}+1+\tilde \rho\geq0.
\end{equation}
Notice that $\tilde \gamma(z)$ is the maximum of the feasible set
$[0, 1-z]$ and the derivative of $f_k$ with respect to $\gamma$ at
$(\tilde{\delta}(z), \tilde \gamma(z))$ is positive. In addition,
$\tilde \delta(z)$ is in the interior of the feasible set $[0,z]$
and the derivative of $f_k$ with respect to $\delta$ at
$(\tilde{\delta}(z), \tilde \gamma(z))$ is zero. Since $f_k$ is
concave, any feasible change in $(\tilde \gamma(z),\tilde
\delta(z))$ will decrease the value of the function. Hence,
$(T_\mu h_k)(z) = (T h_k)(z)$ for $z \in [b_3, b_4]$.  The
situation for $z \in [b_1,b_2]$ is completely analogous.
\hfill\QED

{\it Proof of }{\ref{l_hk}.4}: From Propositions { \ref{l_hk}.1}-{
\ref{l_hk}.3}, it follows that the maximum over $z$ of $h_k(z)$ is
attained at $z=1/2$ and $h_k(1/2) = 1$ for all $k$. Further more
because of concavity and symmetry the minimumm of $h_k(z)$ is
attained at $z=0$ and $z=1$. Hence it is enough to show that
$h_k(0)$ is uniformly bounded from below for all $k$.

For $z=0$ let us consider the action $\gamma=\frac{1-b_2}{1-b_2}$
and $\delta=0$ and for $b_1\leq z\leq b_4$ the action $\tilde
\gamma(z), \tilde \delta(z)$. Now let us prove that under this
policy $h_k(0)$ that is less or equal the optimal value is
uniformly bounded.

Under this policy $h_{k+1}(0) = (T h_k)(0) - \tilde{\rho}$ becomes
\begin{equation}\label{e_iter}
h_k(0)=c+\alpha h_{k-1}(0)+(1-\alpha)1-\tilde \rho
\end{equation}
where $c$ and $\alpha$ are constant:
$c=H\left(\frac{1}{1+b_2}\right)+b_2-1$ ,
$\alpha=\frac{1-b_2}{2}$.

 Iterating the equation (\ref{e_iter})
$k-1$ time we get

\begin{equation}
h_k(0)=(c+1-\alpha-\tilde \rho)\sum_{i=0}^{k-1}\alpha^i+\alpha^k
h_{0}(0).
\end{equation}
Since $\alpha<1$, $h_k(0)$ is uniformly bounded for all $k$.
\hfill \QED

{\it Proof of }{\ref{l_hk}.5}: By Proposition  \ref{l_hk}.1 $h_k$
is concave for each $k$ and by Proposition \ref{l_hk}.3 $h_k(z) =
\tilde{h}(z)$ for $z \in [b_1,b_4]$. Since $h_0$ is the pointwise
maximum of functions satisfying this condition, we must have $h_0
\geq h_1$. It is easy to see that $T$ is a monotonic operator.  As
such, $h_k \geq h_{k+1}$ for all $k$. Proposition \ref{l_hk}.4
establishes that the sequence is bounded below, and therefore it
converges pointwise. \hfill \QED

{\it Proof of }{\ref{l_hk}.6}: By Proposition \ref{l_hk}.1, each
$h_k$ is concave and continuous.  Further, by Proposition
\ref{l_hk}.5, the sequence has a pointwise limit $h^*$ which is
concave. Concavity of $h^*$ implies continuity \cite[Theorem
10.1]{Rockafeller1970} over $(0,1)$. Let $h^\dagger$ be the
continuous extension of $h^*$ from $(0,1)$ to $[0,1]$.  Since
$h^*$ is concave, $h^\dagger \geq h^*$.

By Proposition { \ref{l_hk}.5} , $h_k \geq h^*$. It follows from
continuity of $h_k$ that $h_k \geq h^\dagger$. Hence, $h^*(z) =
\lim_k h_k(z) \geq h^\dagger(z)$ for $z \in [0,1]$.  Recalling
that $h^* \leq h^\dagger$, we have $h^* = h^\dagger$.

Since the iterates $h_k$ are continuous and monotonically
nonincreasing and their pointwise limit $h^*$ is continuous, $h_k$
converges uniformly by Dini's Theorem \cite{Marsden_analysis}.
\hfill \QED

The following theorem verifies our conjectures.
\begin{theorem}
The function $h^*$ and scalar $\tilde{\rho}$ satisfy $\tilde{\rho}
{\bf 1} + h^* = Th^*$. Further, $\tilde{\rho}$ is the optimal
average reward and there is an optimal policy that takes actions
$\delta_t = \tilde{\delta}(z_{t-1})$ and $\gamma_t =
\tilde{\gamma}(z_{t-1})$ whenever $z_{t-1} \in [b_1,b_4]$.
\end{theorem}
\proof  Since the sequence $h_{k+1} = T h_k - \tilde{\rho} {\bf
1}$ converges uniformly and $T$ is sup-norm continuous, $h^* = T
h^* - \tilde{\rho} {\bf 1}$. It follows from Theorem
\ref{th:Bellman} that $\tilde{\rho}$ is the optimal average
reward. Together with Proposition \ref{l_hk}.3, this implies
existence of an optimal policy that takes actions $\delta_t =
\tilde{\delta}(z_{t-1})$ and $\gamma_t = \tilde{\gamma}(z_{t-1})$
whenever $z_{t-1} \in [b_1,b_4]$. \hfill \QED

\section{A Capacity-Achieving Scheme}\label{s_an_achievale_scheme}
In this section we describe a simple encoder and decoder pair that
provides error-free communication through the trapdoor channel
with feedback and known initial state.  We then show that the
rates achievable with this encoding scheme are arbitrarily close
to capacity.

It will be helpful to discuss the input and output of the channel
in different terms. Recall that the state of the channel is known
to the transmitter because it is a deterministic function of the
previous state, input, and output, and the initial state is known.
Let the input action, $\tilde{x}$, be one of the following:
$$\tilde{x} = \left\{ \begin{tabular}{ll} 0, & \mbox{input ball is same as state} \\ 1, & \mbox{input ball is opposite of state} \end{tabular} \right.$$
Also let the output be recorded differentially as,
$$\tilde{y} = \left\{ \begin{tabular}{ll} 0, & \mbox{received ball is same as previous} \\ 1, & \mbox{received ball is opposite of previous} \end{tabular} \right.$$
where $\tilde{y}_1$ is undefined and irrelevant for our scheme.

\subsection{Encode/Decode Scheme}

{\it Encoding.}  Each message is mapped to a unique binary
sequence of $N$ actions, $\tilde{x}_n$, that ends with 0 and has
no occurrences of two 1's in a row.  The input to the channel is
derived from the action and the state as, $x_k = \tilde{x}_k
\oplus s_{k-1}$.
\\

\noindent {\it Decoding.}  The channel outputs are recorded
differentially as, $\tilde{y}_k = y_k \oplus y_{k-1}$, for
$k=2,...,N$.  Decoding of the action sequence is accomplished in
reverse order, beginning with $\tilde{x}_N = 0$ by construction.

\begin{lemma}
\label{l_decode_scheme} If $\tilde{x}_{k+1}$ is known to the
decoder, $\tilde{x}_k$ can be correctly decoded.
\end{lemma}

\begin{proof}
Table \ref{t_decode_rules} shows how to decode $\tilde{x}_k$ from
$\tilde{x}_{k+1}$ and $\tilde{y}_{k+1}$.

\begin{table}[h]
\caption{Decoding the input from the next output and input.}
\centering \label{t_decode_rules}
\begin{tabular}{|l|c|c||c|}
\hline
& $\tilde{y}_{k+1} $ & $\tilde{x}_{k+1}$ & $\tilde{x}_k$ \\
\hline
Case 1 & 0 & - & 0 \\
Case 2 & - & 1 & 0 \\
Case 3 & 1 & 0 & 1 \\
\hline
\end{tabular}
\end{table}

{\it Proof of case 1.} Assume that $\tilde{x}_k = 1$. At time $k$,
just before the output is received, there are balls of both types
in the channel. By symmetry we can assume that the ball that exits
is labeled `0.' Therefore, the ball labeled `1' remains in the
channel.  According to the encoding scheme, $\tilde{x}_{k+1} = 0$
because repeated 1's are not allowed, which means the input to the
channel at time $k$ is labeled `1.'  It is clear that the ball
that comes out of the channel at time $k+1$ must be labeled `1.'
This leads to the contradiction, $\tilde{y}_{k+1} = 1$.

{\it Proof of case 2.} By construction there are never two 1's in
a row.

{\it Proof of case 3.} Assume that $\tilde{x}_k = 0$. The balls
that enter the channel both at times $k$ and $k+1$ are the same
type as the ball that is in the channel, therefore that same type
of ball must come out each of the two times.  This leads to the
contradiction, $\tilde{y}_{k+1} = 0$.

\end{proof}

{\it Decoding example.} Table \ref{t_decode_example} shows an
example of decoding a sequence of actions for $N=10$.

\begin{table}[h]
\caption{Decoding Example} \centering \label{t_decode_example}
\begin{tabular}{l|r|l}
\hline Variable & Value & Reason \\
\hline
\hline $y_n$ & 1011010001 & Channel output\\
\hline $\tilde{y}_n$ & *110111001 & Differential output\\
\hline $\tilde{x}_n$ & 0 & Given \\
& 10 & Case 3 \\
& 010 & Case 1 or 2 \\
& 0010 & Case 1 \\
& 10010 & Case 3 \\
& 010010 & Case 2 \\
& 1010010 & Case 3 \\
& 01010010 & Case 1 or 2 \\
& 101010010 & Case 3 \\
& 0101010010 & Case 2 \\
\end{tabular}
\end{table}

\subsection{Rate}

Under this encoding scheme, the number of admissible unique action
sequences is the number of binary sequences of length $N-1$
without any repeating 1's.  This is known to be exponentially
equivalent to $\phi^{N-1}$, where $\phi$ is the golden ratio (see
question 2 in section \ref{s_conjecture}).  Since $\lim_{N \to
\infty} \frac{N-1}{N} \log \phi = \log \phi$, rates arbitrary
close to $\log \phi$ are achievable.

\subsection{Remarks}

{\it Early decoding.}  Decoding can often begin before the entire
block is received.  Table \ref{t_decode_rules} shows us that we
can decode $\tilde{x}_k$ without knowledge of $\tilde{x}_{k+1}$
for any $k$ such that $\tilde{y}_{k+1} = 0$.  Decoding can begin
from any such point and work backward.
\\

\noindent {\it Preparing the channel.}  This communication scheme
can still be implemented even if the initial state of the channel
is not known as long as some channel uses are expended to prepare
the channel for communication.  The repeating sequence 010101...
can be used to flush the channel until the state becomes evident.
As soon as the output of the channel is different from the input,
both the transmitter (through feedback) and the receiver know that
the state is the previous input.  At that point, zero-error
communication can begin as described above.

This flushing method requires a random and unbounded number of
channel uses.  However, it only needs to be performed once after
which multiple blocks of communication can be accomplished.  The
expected number of required channel uses is easily found to be
3.5, since the number of uses is geometrically distributed when
conditioned on the initial state.
\\

\noindent {\it Permuting relay channel similarity.}  The permuting
relay channel described in \cite{Ahl_kaspi87} has the same
capacity as the trapdoor channel with feedback.  A connection can
be made using the achievable scheme described in this section.

The permuting relay channel supposes that the transmitter chooses
an input distribution to the channel that is independent of the
message to be sent.  The transmitter lives inside the trapdoor
channel and chooses which of the two balls will be released to the
receiver in order to send the message.  Without proof here, let us
assume that the deterministic input 010101... is optimal.  Now we
count how many distinguishable outputs are possible.

It is helpful to view this as a permutation channel as described
in section \ref{s_channel_model_and_pre} where the permuting is
not done randomly but deliberately.  Notice that for this input
sequence, after each time that a pair of different numbers is
permuted, the next pair of numbers will be the same, and the
associated action will have no consequence.  Therefore, the number
of distinguishable permutations can be easily shown to be related
to the number of unique binary sequences without two 1's in a row.
\\

\noindent {\it Three channels have same feedback capacity.}  The
achievable scheme in this section allows zero-error communication.
Therefore, this scheme could also be used to communicate with
feedback through the permuting jammer channel from
\cite{Ahl_kaspi87}, which assumes that the trapdoor channel
behavior is not random but is the worst possible to make
communication difficult.

In the permuting relay channel \cite{Ahl_kaspi87}, all information
(input and output) is available to the transmitter, so feedback is
irrelevant.  Thus we find that the feedback capacity (with known
initial state) is the same for the trapdoor, permuting jammer, and
permuting relay channels.
\\

\noindent {\it Constrained coding.} The capacity-achieving scheme
requires uniquely mapping a message to a sequence with the
constraint of having no two 1's in a row. A practical way of
accomplishing this can be done by using a technique called {\it
enumeration} \cite{Cover73}. The technique translates the message
into codewords and vice versa by invoking an algorithmic procedure
rather then using a lookup table. Vast literature on coding a
source word into a constrained sequence
 can be found in \cite{Marcus98} and \cite{Immink04}.
\\

\section{Conclusion and Further Work}\label{s_conclusion}
This paper gives an information theory formulation for the
feedback capacity of a strongly connected unifilar finite state
channel and it shows that the feedback capacity expression can be
formulated as an average-reward dynamic program. For the trapdoor
channel, we were able to solve explicitly the dynamic programming
problem and to show that the capacity of the channel is the log of
the golden ratio. Furthermore, we were able to find a simple
encoding/decoding scheme that achieves this capacity.

There are several directions in which this work can be extended.
\begin{itemize}
\item {\it Generalization:} Extend the trapdoor channel
definition. It is possible to add parameters to the channel and
make it more general. For instance, there could be a parameter
that determines which ball from the two has the higher probability
of being the output of the channel.  Other parameters might
include the number of balls that can be in the channel at the same
time or the number of different types of balls that are used.
These tie in nicely with viewing the trapdoor channel as a
chemical channel.  \item {\it Unifilar FSC Problems:} Find
strongly connected unifilar FSC's that can be solved, similar to
the way we solved the trapdoor channel. \item {\it Dynamic
Programming:} Classify a family of average-reward dynamic programs
that have analytic solutions.
\end{itemize}

\section*{Acknowledgment}
The authors would like to thank Tom Cover, who introduced the
trapdoor channel to H. Permuter and P. Cuff and asked them the two
questions that appear in subsection \ref{s_conjecture}, which
eventually led to the solution of the dynamic programming and to
the simple scheme that achieves the feedback capacity.

\bibliographystyle{unsrt}
\bibliography{C:/haim/mydoc/GE-channel/ref}

%\appendices
\appendix
%\section{Proof of Lemma \ref{l_concavity}}\label{a_l_concavity}
%\begin{proof}
{\it Proof of Lemma \ref{l_concavity}:} For any $z_1, z_2 \in
[0,1]$ and $\theta \in (0,1)$,
\begin{eqnarray}
\psi(\theta z_1+ (1-\theta) z_2)&=& \sup_{\delta \in [0, \theta
z_1+(1-\theta) z_2]} \sup_{\gamma \in [0, 1-(\theta z_1+(1-\theta)
z_2)]}
\zeta(\delta,\gamma)\nonumber \\
&\stackrel{}{=}& \sup_{\delta_1 \in [0, \theta z_1]}
\sup_{\delta_2 \in [0, (1-\theta) z_2]} \sup_{\gamma_1 \in [0,
\theta (1- z_1)]} \sup_{\gamma_2 \in [0,(1-\theta)(1-z_2)]}
\zeta(\delta_1+\delta_2,\gamma_1+\gamma_2)
\nonumber \\
&\stackrel{(a)}{=}& \sup_{\delta'_1 \in [0,z_1]} \sup_{\delta'_2
\in [0,z_2]} \sup_{\gamma'_1 \in [0, 1-z_1]} \sup_{\gamma'_2 \in
[0,1-z_2]} \zeta(\theta \delta'_1+(1-\theta) \delta'_2, \theta
\gamma'_1+(1-\theta)\gamma'_2)\nonumber \\
&\stackrel{(b)}{\geq}& \sup_{\delta'_1 \in [0,z_1]} \sup_{
\delta'_2 \in [0,z_2]} \sup_{\gamma'_1 \in [0,1- z_1]}
\sup_{\gamma'_2 \in [0,1-z_2]}
\theta \zeta(\delta'_1,\gamma'_1)+(1-\theta)\zeta(\delta'_2,\gamma'_2)\nonumber \\
\nonumber \\
&\stackrel{}{=}& \sup_{\delta'_1 \in [0,z_1]} \sup_{\gamma'_1 \in
[0,1- z_1]} \theta \zeta(\delta'_1,\gamma'_1) + \sup_{\delta'_2
\in [0,z_2]} \sup_{\gamma'_2 \in [0,1-z_2]}
(1-\theta) \zeta(\delta'_2,\gamma'_2) \nonumber \\
&=& \theta \psi(z_1)+(1-\theta) \psi(z_2).
 \end{eqnarray}
Step (a) is a change of variable $(\theta
\delta'_1=\delta_1,(1-\theta)\delta'_2=\delta_2,\theta
\gamma'_1=\gamma_1,(1-\theta)\gamma'_2=\gamma_2)$.  Step (b) is
due to concavity of $\zeta$.
%\end{proof}
\hfill \QED
\end{document}